\documentclass[a4paper, fleqn, oneside]{article}
\usepackage{amsmath}
\usepackage{amsfonts}
\usepackage{amsthm}

\DeclareMathOperator{\supp}{supp}
\DeclareMathOperator{\diverg}{div}
\DeclareMathOperator{\id}{id}

\DeclareMathOperator{\sgn}{sgn}

\newtheorem{lemma}{Lemma}[section]
\newtheorem{theorem}{Theorem}[section]
\newtheorem{proposition}{Proposition}[section]

\newtheorem{corollary}{Corollary}[section]
\newtheorem{definition}{Definition}[section]

{\catcode `\@=11 \global\let\AddToReset=\@addtoreset}
\AddToReset{equation}{section}
\newcommand{\fer}[1]{(\ref{#1})} 
\newcommand{\be}{\begin{equation}}
\newcommand{\ee}{\end{equation}}
\newcommand{\Jac}{J}
\newcommand{\eps}{\varepsilon}
\newcommand{\ep}{\varepsilon}
\newcommand{\R}{\mathbb R}
\newcommand{\A}{\mathcal A}
\renewcommand{\S}{\mathcal S}

\begin{document}
\thispagestyle{empty}
\begin{center}
\fontsize{22}{22}
\selectfont
\vskip 3cm 
\textbf{Wigner Functions versus WKB-Methods in Multivalued Geometrical Optics}
\end{center}
\vskip 2 cm
\vskip 2 cm
\begin{center}
\fontsize{12}{12}
\selectfont
Christof Sparber$^\star $, Peter A. Markowich$^\dagger $, Norbert J.
Mauser$^\ddagger $\\
\vskip 0.5 cm
Inst. f. Mathematik, Univ. Wien, \\
Boltzmanngasse 9, A-1090 Vienna,\\
Austria\\
\end{center}
\vskip 1 cm
\begin{center}
{\textbf{Abstract}}
\end{center}
\emph{We consider the Cauchy problem for a class of scalar linear dispersive
equations with rapidly oscillating initial data. The problem of the high-frequency asymptotics of such
models is reviewed, in particular we highlight the difficulties in crossing caustics when using
(time-dependent) WKB-methods. 
Using Wigner measures we present an alternative approach to such asymptotic problems.
We first discuss the connection of the naive WKB solutions to transport equations of
Liouville type (with mono-kinetic solutions) in the  prebreaking regime. Further we show how the Wigner measure
approach can be used to analyze high-frequency limits in the post-breaking regime, in comparson with 
the traditional Fourier integral operator method.  
Finally we present some illustrating examples.$^0$}
\vskip2cm
\footnotetext{This work was supported by the
Austrian \emph{START award} (FWF Y-137-TEC) of N.J.M., 
the \emph{"Wittgenstein Award"} of P.A.M. 
and by the FWF Wissenschaftskolleg "Differential Equations".\\
$^\ast $ e-mail: christof.sparber@univie.ac.at\\
$^\dagger $ e-mail: peter.markowich@univie.ac.at or http://mailbox.univie.ac.at/peter.markowich\\
$^\ddagger $ e-mail: mauser@courant.nyu.edu} 
\clearpage

\section{Introduction}
\fontsize{11}{12}
\selectfont
In this  paper we consider a class of scalar IVP's for linear dispersive
equations 
with 
fast temporal and spatial scales  subject to highly oscillating initial data. 
The Cauchy problem of the Schr\"odinger equation serves as a typical
example 
\begin{eqnarray}
\label{intsch}i\varepsilon \partial _t\psi^\varepsilon &=& -\frac{\varepsilon
^2}{2}\Delta \psi^\varepsilon + 
V(x)\psi ^\varepsilon =0,\quad x\in\mathbb R^d, t\in \mathbb R\\
\psi ^\varepsilon (x,0) &=& A_I(x)e^{i S_I(x)/\varepsilon}, \quad
x\in\mathbb R^d
\end{eqnarray}
where $\varepsilon \sim \hbar$ (the scaled Planck's constant). The small
parameter $\varepsilon $ 
represents the fast space and time scales introduced in (1.1), as well as the
typical wave length of oscillations 
of the initial data. We are interested in the high frequency limit of these
equations, which is usually referred 
to as \emph{"geometrical-optics"}. 
In the special case of the Schr\"odinger equation with vanishing Planck's
constant this is precisely the 
\emph{"(semi-)classical limit".} 
It is well known that the considered equations propagate oscillations of wave
lengths $\varepsilon $ 
which inhibit $\psi ^\varepsilon $ from converging strongly in a suitable sense.\\
Thus the short-wavelength-asymptotics $\varepsilon \rightarrow 0$ is by no means
straightforward, 
in particular since the physical quantities of interest (observables) are
quadratic in $\psi ^\varepsilon $.\\
The usual way to tackle the problem is the geometrical optics -  or
\emph{WKB-Ansatz} (Wentzel-Kramers-Brillouin, 
see \cite{Ke}), which consists of representing the solution $\psi^\varepsilon $
in the form
\begin{equation}
\label{ans}\psi ^\varepsilon (x,t)\simeq  A^\varepsilon  (x,t)
\exp\left({\frac{i}{\varepsilon }S(x,t)}\right)
\end{equation}
where $A^\varepsilon$ and $S$ are realvalued, $A^\varepsilon \geq 0$ 
and in general $A^\varepsilon \simeq  A+\varepsilon A_1+\varepsilon ^2A_2+\dots$.\\
Then after inserting the above representation into the equation and by
considering, 
as a first approximation, only the lowest order terms, one finds that:
\begin{itemize}
\item the phase $S$ is a solution of a nonlinear first order equation of
\emph{Hamilton-Jacobi type}
\item the (zeroth order) amplitude satisfies a linear first order PDE (called
\emph{transport equation}) 
that can be brought into the form of a \emph{conservation law} for the
\emph{energy density} $n=A^2$.
\end{itemize}
A severe drawback of this method should be noted. The obtained nonlinear
equations do not have global, 
i.e. for all $t\in\mathbb R$, smooth solutions (except for some special initial
data). 
In other words the system in general develops singularities in some finite time
$t_c$ (\emph{"break time"}). 
The formal expansion method clearly can only be justified for smooth, i.e.
sufficiently often differentiable, 
functions $S$ and $A^\varepsilon  $ and thus the Ansatz (\ref{ans}) breaks down
at points where the first 
singularities occur. These singularities are called \emph{focal points}, or more
generally \emph{caustics}, 
since, as we will see, the energy of the wave becomes infinite there. \\
A alternative point of view on this problem is given by considering the so called 
\emph{Quantum Hydrodynamic System}, which is 
obtained by plugging into equation (\ref{intsch}) the ansatz (\ref{ans}), with $S=S^\varepsilon $. Defining  
$n^\varepsilon :={(A^\varepsilon) }^2$ and $j^\varepsilon :=n^\varepsilon \nabla S^\varepsilon $, one gets 
(after seperating real and imaginary parts) 
\begin{eqnarray}
\label{qhd1}\partial _t n^\varepsilon +  \diverg j^\varepsilon  &=&0,\\
\label{qhd2}\partial _t j^\varepsilon + \diverg \left(\frac{j^\varepsilon \otimes
j^\varepsilon}{n^\varepsilon } \right)
+ n^\varepsilon  \nabla_x V &=
& \frac{\varepsilon ^2}{2}n^\varepsilon \nabla \left(\frac{1}{\sqrt{ n^\varepsilon}}\Delta
\sqrt{n^\varepsilon}\right).
\end{eqnarray}
This system is exact, i.e. equivalent to the Schr\"odinger equation (\ref{intsch}), c.f. \cite{GaMa} 
and well posed for all $t\in\mathbb R$ due to the third order dispersive regularization term. 
However for $\varepsilon =0$, where the system simplifies to the zero temperature Euler equations, 
singularities appear and the equations cannot be used after them 
to describe the propagation of the energy density $n^\varepsilon $ in the high frequency limit. 
(This approach is used in particular in one space variable for the non linear Schr\"odinger equation, i.e. 
$V^\varepsilon =|\psi ^\varepsilon |^2$,  see e.g. \cite{LaLe}.)  \\
A natural alternative to the standard WKB-method is seeking \emph{multivalued
phases} 
corresponding to crossing waves. This means that in general for every fixed
$(x,t)\in\mathbb R^d\times \mathbb R$, 
which is not on the caustic, one tries to construct a (maybe infinite) set of
phase 
functions $\{S_i(x,t)\}$, $i\in I\subseteq \mathbb N$, each of which is a
solution of the Hamilton-Jacobi equation 
in a neighborhood of $(x,t)$. 
This set is referred to as the \emph{multivalued solution of the Hamilton-Jacobi
equation} 
and it induces the multivalued solution $\{n_i(x,t)\}$ of the conservation law
for the energy density.\\ 
\noindent Historically this problem was studied by P. Lax, D. Ludwig, V. Maslov,
J. Duistermaat and others 
(see \cite{Du}, \cite{Kr}, \cite{La}, \cite{Lu}, \cite{Mas1}), who showed that 
\emph{Fourier-Integral operators} and \emph{Lagrangian manifolds} in \emph{phase
space} 
provide a uniform description of of the behavior of $\psi^\varepsilon $. 
(For applications in semi-classical quantum mechanics see the expository article of Robert \cite{Ro}.) 
The qualitative study of the multivalued solution is accomplished using
geometrical techniques 
of singularity theory and contact geometry (see e.g. \cite{Ar}, \cite{AVG}, \cite{Du}, \cite{GuSt}). \\
A considerable amount of work has been done in recent years on constructing
numerically 
the multivalued phase function (see, e.g., \cite{Be1}, \cite{Be2}, \cite{BKM},
\cite{Ru}, \cite{JiLi}).  
We will not cover the arising numerical questions in this paper, instead we
refer the 
interested reader to these references.\\
In the last decade the use of \emph{Wigner functions} and \emph{Wigner measures}
has drawn increasing interest, in particular its application to the
semi-classical 
limit of Schr\"odinger equations (\cite{LiPa}, \cite{MaMa}, \cite{MMP},
\cite{MPP}) 
and the homogenization of energy densities of dispersive equations 
(e.g. \cite{BCKP}, \cite{GaMa}, \cite{Ge}, \cite{GMMP}), mostly by groups in
Europe.
Independently, groups in the US used Wigner functions, too, with some emphasis
on waves in 
random media and applied problems, e.g. \cite{BKPR}, \cite{PaRh}.\\
The Wigner transformation provides a phase space description of the equations of
the problem, 
which is extremely useful for the asymptotics since it "unfolds" the
caustics (de-projection in phase space). 
Another advantage is that the high frequency limit, using Wigner functions,
needs much lower regularity 
assumptions on $A_I$ and $S_I$ than the (generalized) WKB-method. 
This is not only of purely academic interest since very often in concrete
physical models $C_0^\infty $ 
initial data are simply not available. The price paid for the analytical
convenience lies in the doubling 
of the dimension, i.e. the Wigner function is defined on $\mathbb R^{2d}$.\\
This work studies the connection between the WKB-method and the Wigner
transformation or, in other words,  
represents an alternative approach to WKB-asymptotics. It is organized as
follows:\begin{itemize}
\item Section 2 is devoted to the setting of the problem. 
There we also give a short review of the traditional (naive) WKB-method 
and its generalization using FIO's.
\item In section 3 we present the main theorems on Wigner transforms and show
how they can be used to obtain a semi-classical phase-space description.
\item This limiting phase-space regime is analyzed in section 4, which is the
most important part of this paper.
\item Examples are studied in section 5 to illustrate the results of the
foregoing section.
\item Finally in the Appendix in section 7 we give an example with non-global
Hamiltonian flow and comment on the arising fluid type equations, which generalize the zero temperature 
Euler equations.
\end{itemize}


\section{Setting of the problem and  the WKB-method}
\subsection{The model equation}
We consider the following initial value problem (\emph{generalized linear
dispersive model}) for an 
anti-selfadjoint scalar pseudo-differential operator
\begin {equation} \label{disp}
\varepsilon \partial _t\psi^\varepsilon  + iH^W(x,\varepsilon D)\psi^\varepsilon
= 0,
\quad x\in \mathbb{R}^d, t\in \mathbb{R}
\end{equation}
subject to the highly oscillatory (WKB) initial data
\begin {equation} \label{ho}
\psi^\varepsilon (x,0)=\sqrt{n_I(x)}e^{i S_I(x)/\varepsilon}, \quad x\in\mathbb
R^d 
\end{equation}
where $\psi^\varepsilon (t,x)$ is a scalar $L^2$-function on $\mathbb R^d$,
$D:=
-i\nabla_x$, $\varepsilon\in(0, \varepsilon _0] $ 
and $H(x,\varepsilon D)^W$ is the \emph{Weyl-operator} 
associated to its symbol $H(x,\varepsilon \xi )$ by \emph{Weyl's quantization
rule:}
\begin {definition}
\begin{equation} \label{Weylop-def}
H^W(x,\varepsilon D) \varphi (x):=
\frac{1}{(2\pi )^d}\int_{\mathbb R^d}\int_{\mathbb R^d}H\left(
\frac{x+y}{2},\varepsilon \xi \right) \varphi (y)
e^{i(x-y)\xi }d\xi dy
\end{equation}
\end{definition}
\noindent The convenience of the Weyl-calculus lies in the fact that a scalar Weyl-operator
is formally selfadjoint iff it has a real-valued symbol.\\
We have chosen this particular form of pseudo differential-calculus in order to be
consistent with the usual 
framework of the Wigner-functions introduced in \cite{GMMP}. Note that
in case $H$ is a sum of 
separate terms in $x$ and $\xi$, the "Weyl symbol" and the "left
symbol"of the classical Fourier multiplier coincide. \\

\noindent \textbf{Remarks.} \begin{itemize}
\item The general framework of pseudo-differential operators allows us to include also 
non-local Hamiltonians, like the one appearing in example (iii) below, in our discussion.
\item Further note that the time and spatial scales of (\ref{disp}) are
"fast", since the small parameter 
$\varepsilon $ multiplies the time and spatial derivatives. 
\end{itemize}

\noindent We shall use in this text the following definition of the 
\emph{Fourier transform} $\mathcal F: \mathcal S(\mathbb R^d)\longrightarrow \mathcal S(\mathbb R^d) $:
\begin{equation} \label{FTdef}
\hat f(\xi ):=(\mathcal F_{x\rightarrow \xi }f)(\xi
):=\int_{\mathbb{R}^d}f(x)e^{-ix\xi }dx,
\end{equation}
with the usual extension, by duality, to a mapping from $\mathcal S'$ to
$\mathcal S'$. \\
In (\ref{ho}) the amplitude is written in this particular form to match the
following definition: 
\begin {definition} The energy-density of the solution of (\ref{disp}) is
defined by
\begin{equation}
\label{en}n^\varepsilon (x,t):=|\psi ^\varepsilon (x,t)|^2 .
\end{equation}
\end {definition}

\noindent We assume on the Weyl operator $H^W$ and on its symbol $H$:\\

\noindent \textbf{Assumption (A1)} \\

\noindent (A1)(i) $\exists$ $\sigma \in \mathbb{R}: H\in S^\sigma
(\mathbb{R}^d)$ \emph{uniformly for} $\varepsilon \in(0,\varepsilon _0]$. \\
(A1)(ii) $\exists$ \emph{a unique self adjoint extension of} $iH^W(\cdot,
\varepsilon D)$ \emph{on} $L^2(\mathbb{R}^d)$.\\ 

\noindent By abuse of notation, we denote the unique s.a. extension by
$iH^W$ too.\\

\noindent The hypothesis (A1)(i) means (see also \cite{Ho}) that for all $\alpha, \beta
\in \mathbb N_0$, there exists $C_{\alpha,\beta  }\geq 0$ , s.t. for all  $l,k
\in\{1..m\}$ and for all $\varepsilon \in(0,\varepsilon _0]$ it holds
\begin{equation*}
| \frac{\partial^{\alpha+\beta }}{\partial x_k^\alpha \partial \xi _l^\beta }
H(x,\xi )|\leq C_{\alpha,\beta}(1+\mid \xi \mid )^{\sigma-\beta}, \quad\forall
(x,\xi)\in\mathbb R^d\times \mathbb R^d.
\end{equation*}
\noindent In particular this implies that the Sobolev space $H^\sigma (\mathbb
R^d)$ lies in the domain of the operator $H^W$. \\
Furthermore this type of hypothesis are made in order to extend the rule of composition 
from differential operators to pseudo-differential operators modulo lower order 
terms in $\varepsilon $ (for details see e.g. \cite{Ho}).\\ 

\noindent\textbf{Remark.} We remark that the regularity assumptions on the symbol
$H$ are largely used 
for convenience and taken from \cite{GMMP} in order to use certain results,
which were established 
there. 
They can be significantly weakened, for example: If $H=|\xi |^2/2+V(x)$ 
with $V$ bounded below and $V\in C^{1,1}$ (cf \cite{LiPa}), 
all results using Wigner transforms remain valid (see  \cite{GaMa}). \\

\noindent On the initial data we impose:\\

\noindent \textbf{Assumption (A2)} 
\begin{equation} \label{A2}
n_I\in L^1(\mathbb R^d) \: , \: n_I\geq 0 \mbox{ a.e. on $\mathbb R^d$} 
\quad \mbox{and } S_I \in W_{loc}^{1,1}(\mathbb R^d).
\end{equation}
\noindent Note that due to the low regularity assumed in (A2) a traditional
WKB-expansion method would not be possible here ! This is one of the advantages
of the Wigner formalism.
\begin{lemma} Let $\psi_I ^\varepsilon$ satisfy (A2) and assume (A1), then there
exist a unique mild solution $\psi^\varepsilon(t) \in C(\mathbb R_t; L^2(\mathbb
R^d))$ of (\ref{disp}), and its energy-density satisfies
\begin {equation}
\label{cons}{\| n^\varepsilon (t)\| } _{1}=
{\| n_I^\varepsilon \| } _{1}\quad\forall \ t\in \mathbb R,
\end{equation}
where $\| \cdot \| _1$ denotes the $L^1$ norm.
\end{lemma} 
\begin{proof} Having in mind (A1)(ii) the assertion is a simple consequence of
Stone's famous theorem (see e.g. \cite{ReSi}).
\end{proof}
\noindent Some particular examples for equation (\ref{disp}) are:\\

\noindent\textbf{Examples:}\\

\noindent (i) The \emph{Schr\"odinger equation}
\begin{equation} \label{schro}
\varepsilon \partial _t\psi^\varepsilon -i \frac{\varepsilon ^2 }{2} \Delta 
\psi^\varepsilon  +iV(x)\psi^\varepsilon=0, 
\end{equation}
where $\varepsilon \sim \hbar $. Here $H(x,\xi )=|\xi|^2/2+V(x)$.\\

\noindent (ii) The 1-d \emph{Airy equation} (or linearized KdV equation)
\begin{equation}
\label{airy}\varepsilon \partial _t\psi^\varepsilon -\frac{\varepsilon
^3}{3}\psi _{xxx}^\varepsilon =0, 
\end{equation} 
where $H(x,\xi ) = \xi ^3/3$. Here $\varepsilon $ denotes the
"physical" dispersion-parameter.\\

\noindent (iii) The \emph{spinless Bethe-Salpeter equation} (or
"relativistic Schr\"odinger equation")
\begin{equation}
\label{bseq}\varepsilon \partial _t\psi ^\varepsilon
-i\left(\sqrt{-\frac{\varepsilon ^2}{2}\Delta  +1}+V(x)\right)\psi ^\varepsilon
=0 .
\end{equation}
Again $\varepsilon \sim \hbar $ and we have $H(x,\xi )=\sqrt{|\xi
|^2/2+1}+V(x)$. Note that in this example $H^W$ is a "true"
pseudo-differential operator (i.e. the Weyl-symbol is not polynomial in $\xi
$).\\

\noindent (iv) Another example of a "true" pseudo-differential
equation assumes Hamiltonians of the form $H=a(x)|\xi |$, i.e. we have (in
Weyl-quantized form)
\begin{equation}
\label{pseu}\varepsilon \partial _t\psi ^\varepsilon +i\varepsilon |D
_y|(a(\frac{x+y}{2})\psi ^\varepsilon(y))\big|_{y=x} =0 .
\end{equation}
In the constant coefficient case $a(x)\equiv 1$, equations of this type can be
traced back to the wave equation $u^\varepsilon _{tt}-\Delta u^\varepsilon =0$
by noting that, the quantities
\begin{equation*}
\psi ^\varepsilon_{\pm }(\xi ,t)=\partial _t u^\varepsilon \pm i|D|u^\varepsilon
\end{equation*}
satisfy 
\begin{equation*}
\partial _t \psi^\varepsilon  _{\pm}\mp i|D |\psi^\varepsilon _{\pm } =0 .
\end{equation*}
\hfill$\diamondsuit $\\

\noindent We are now interested in the high-frequency limit $\varepsilon
\rightarrow 0$. 
For the sake of completeness we briefly review the traditional (naive) WKB-method in 
the next subsection.


\subsection{The WKB-method}
As stated in the introduction above we make the following ansatz 
\begin{equation} \label{wkbrep}\
\psi ^\varepsilon (x,t) \simeq  
A^\varepsilon  (x,t) \exp\left({\frac{i}{\varepsilon }S(x,t)}\right), \quad
x\in\mathbb R^d, t\in\mathbb R
\end{equation}
with $A^\varepsilon \geq 0$ assuming (for the moment) that the phase and the
amplitude are sufficiently smooth, 
and we expand the amplitude in powers of $\varepsilon $:
\begin{equation*}
A^\varepsilon \simeq A+\varepsilon A_1+\varepsilon ^2A_2+\dots
\end{equation*}
\noindent We sketch the (formal) WKB-method for the case of a polynomial
Weyl-symbol $H=H(x,\xi)$ with $C^\infty$-coefficients, i.e. 
\begin{equation}
\label{polweyl} H^W(x,\varepsilon D)\varphi(x) = 
\sum_{|k|=0}^m \varepsilon^{|k|} D^k_y\bigg(a_k\Big(\frac{x+y}{2}\Big)
\varphi(y)\bigg)\bigg|_{y=x} .
\end{equation}
Substituting the representation (\ref{wkbrep}) into (\ref{disp}) and collecting
terms appropriately gives
\begin{equation}
H^W(x,\varepsilon D) ( A^\varepsilon  e^{iS/\varepsilon}) =
e^{iS/\varepsilon}\sum_{|k|=0}^m(i\varepsilon )^{|k|} R_k[A^\varepsilon ],
\end{equation}
where $R_k$ acts on $A^\varepsilon $ as a differential operator of order
$|k|\leq m$ 
(with coefficients depending on derivatives of $S,H$). In particular
\begin{eqnarray*}
R_0 [A^\varepsilon ]&=& H(x, \nabla_x S) A^\varepsilon, \\
\label{r}R_1[A^\varepsilon ]&=&\sum_{j=1}^d\frac{\partial H(x,\nabla
_xS)}{\partial \xi_j}\frac{\partial A^\varepsilon }{\partial
x_j}+\frac{1}{2}\sum_{j,k=1}^d\frac{\partial^2 S}{\partial x_j\partial
x_k}\frac{\partial^2 H(x,\nabla _xS)}{\partial \xi_j\partial \xi_k}A^\varepsilon
\\
&+&\frac{1}{2}\sum_{k=1}^d \frac{\partial ^2 H(x, \nabla _xS)}{\partial
y_k\partial \xi _k}A^\varepsilon,
\end{eqnarray*}
where here and in the sequel we denote by $\nabla _y$ the gradient w.r.t. the
position variable, i.e. we consider $y$ as a placeholder for the position
variable $x$: $S=S(y,t)$, $H=H(y,\xi )$ etc.. 
The last term in the expression $R_1$ is obtained due to the fact that we use
Weyl-quantized operators. Note that in the equations above all partial
derivatives of the symbol $H$ w.r.t. $\xi $ are evaluated at $\xi =\nabla _xS$.
Plugging the above computations into (\ref{polweyl}), separating real and
imaginary parts terms we obtain in the lowest orders
\begin{eqnarray*}
\frac{\partial S}{\partial t} + H(x,\nabla _xS)&=&0, \\
\frac{\partial A}{\partial t} + \sum_{j=1}^d\frac{\partial H(x,\nabla
_xS)}{\partial \xi_j}\frac{\partial A}{\partial
x_j}+\frac{1}{2}\sum_{j,k=1}^d\frac{\partial^2 S}{\partial x_j\partial
x_k}\frac{\partial^2 H(x,\nabla _xS)}{\partial \xi_j\partial \xi_k}
A\nonumber\\
+\frac{1}{2}\sum_{k=1}^d \frac{\partial ^2 H(x, \nabla _xS)}{\partial
y_k\partial \xi _k}A&=&0 .
\end{eqnarray*}
Note that the second equation is linear in $A$, it is called the \emph{transport
equation} for the amplitude. 
In terms of $(n, S)\equiv ( A^2, S)$ we obtain the following compact form, 
in the sequel called \emph{WKB-system} associated to equations of type
(\ref{disp}) 
\begin{eqnarray}
\label{wkb1} \partial _tn+\diverg(n\nabla _\xi H(x,\nabla_x S))&=&0,
\quad x\in\mathbb R^d, t\in \mathbb R \\
\label{wkb2} \partial  _tS+H(x, \nabla _xS )&=&0, 
\end{eqnarray}
where the initial data are induced by the initial condition (\ref{ho})
\begin{equation}
n(x,0)=n_I(x), \quad S(x,0)=S_I(x),\quad x\in\mathbb R^d.
\end{equation}
The first equation (\ref{wkb1}) is a \emph{conservation law} for the
energy-density, the second (\ref{wkb2}) a \emph{Hamilton-Jacobi equation} for
the phase.  \\

\noindent To illustrate the method, we derive the (lowest order) WKB-systems for
some particular examples of (\ref{disp}).\\

\noindent\textbf{Examples:}\\

\noindent (i) WKB-system for the Schr\"odinger equation (\ref{schro}):
\begin{eqnarray}
\partial _t n+  \diverg (n\nabla _xS) &=&0,\quad x\in\mathbb R^d, t\in
\mathbb R \\
\partial _tS+\frac{|\nabla _xS|^2}{2}+V(x)&=&0.
\end{eqnarray}
\noindent (ii) WKB-system for the 1-d Airy equation (\ref{airy}):
\begin{eqnarray} \label{Airy1}
\partial _tn+ \partial _x(n(\partial _xS)^2 )&=&0,\quad x, t\in \mathbb R \\
\partial _tS + \frac{(\partial _xS)^3}{3}&=&0.
\end{eqnarray}
\noindent (iii) WKB-system for the (spinless) Bethe-Salpeter equation
(\ref{bseq}):
\begin{eqnarray}
\partial _tn+\diverg\left(\frac{n\nabla _xS}{\sqrt{\frac{|\nabla _xS
|^2}{2}+1}}\right)&=&0,\quad x\in\mathbb R^d, t\in \mathbb R \\
\partial _tS+\sqrt{\frac{\vert \nabla_x S \vert ^2}{2}+1}+V(x)&=&0.
\end{eqnarray}
\noindent (iv) WKB-system for equations of type (\ref{pseu}):
\begin{eqnarray}
\partial _tn+\diverg\left(\frac{na\nabla _xS}{|\nabla _xS
|}\right)&=&0,\quad x\in\mathbb R^d, t\in \mathbb R \\
\label{eiko}\partial _tS+a(x)| \nabla_x S| &=&0.
\end{eqnarray}
The equation (\ref{eiko}) is the time-dependent \emph{eikonal equation} of
geometrical optics. Note that in one spatial dimension, i.e. $d=1$, the above
system de-couples and simplifies to:
\begin{eqnarray}
\partial _tn+ \partial _x(na\sgn(\partial _xS))&=&0,\quad x, t\in \mathbb R
\\
\partial _tS+ a(x)\sgn(\partial _xS) \partial_x S&=&0.
\end{eqnarray}
In this case the WKB-approximation is \emph{exact} if $\partial _xS$ does not
change sign, since then no terms $\sim O( \varepsilon ^2)$ appear in the
expansion when using $A_k=0$  for $k>1$.

\hfill$\diamondsuit $\\

\noindent In general, solving the system (\ref{wkb1}), (\ref{wkb2}) allows an
asymptotic description of the solution of (\ref{disp}) in the pre-breaking
regime, more precisely we have the following proposition, the proof of which is 
classic, see e.g. \cite{La}:
\begin{proposition} Let $\psi ^\varepsilon (t)$ solve (\ref{disp}), (\ref{ho}),
where $H^W$ is of the form (\ref{polweyl}). Assume $\sqrt{n_I}\in\mathcal
S(\mathbb R^d)$ and that $S_I$ is bounded, together with all its derivatives up
to sufficient order. Let $t_{c_1}< 0< t_{c_2}$ be such that a smooth
solution $( n,S)$ of (\ref{wkb1}), (\ref{wkb2}) exists on $\mathbb R^d\times
(t_{c_1}, t_{c_2})$ and define
\begin{equation}
\psi ^\varepsilon _{wkb}(x,t):= \sqrt{n(x,t)}\exp\left(\frac{i}{\varepsilon
}S(x,t)\right) .
\end{equation}
Then, for every finite time-interval $[t_1, t_2]\subset (t_{c_1}, t_{c_2})$ we
have 
\begin{equation}
\sup_{t_{1}<t<t_{2}} {\|  \psi ^\varepsilon (t)-\psi ^\varepsilon
_{wkb}(t)\| }_{2}\leq O(\varepsilon ) ,
\end{equation}
where $\| \cdot \| _2$ denotes the $L^2$ norm.
\end{proposition}
\noindent \textbf{Remark.} 
The proposition can be extended to (systems of ) pseudo-differential
operators, with somewhat weaker conditions on $n$, $S$ and $H$ (see e.g. \cite{Fed}, \cite{Mas1}, \cite{Mas2}).  \\

\noindent Generally one faces the following problem in this approach:\\

\noindent The solution $S$ of the Hamilton-Jacobi equation (\ref{wkb2}) is
obtained, at least locally, by the \emph{method of characteristics}. This means
that, for a fixed $t$, each point $(y,t)\in\mathcal U$ in a (sufficiently small)
neighborhood $\mathcal U\subset \mathbb R_x^d\times \mathbb R_t$ of the initial
manifold $\mathbb R_x^d\times \{t=0\}$, is reached by a unique integral curve
$\hat x(t,x)$, called \emph{ray}, which can be found as described  in the
following.\\
Let us define 
\begin{equation}
\hat \xi (t,x) :=\nabla _{y}S(\hat x(t,x),t).
\end{equation}
It is well known (see for example \cite{Ev}), that the curves $\hat x(t,x),\hat
\xi (t,x)$ solve the IVP
\begin{eqnarray}
\label{hode1}\frac {d{\hat  x}}{dt} &=& \nabla _{\xi} H({\hat  x},{\hat
\xi }), \quad \hat {x}(0,x)= x\\
\label{hode2}\frac {d{\hat  \xi }}{dt} &=& -\nabla _{y} H({\hat 
x},{\hat  \xi }),\quad \hat {\xi }(0,x)= 
\nabla_x S_I(x).
\end{eqnarray}
Having solved this system we obtain the solution of (\ref{wkb2}) by integrating
\begin{equation}
\label{phas}\frac{dS({\hat x},t)}{dt}=\nabla _{\xi} H(\hat x, \hat \xi )\cdot
\hat \xi 
- H(\hat x,\hat \xi ),\quad S(\hat x(0,x),0)=S_I(x).
\end{equation}
For details see again \cite{Ev}, \cite{Fed}. As it is indicated above this
theory is \emph{local}, 
since it only gives the unique solution of (\ref{wkb2}) in a neighborhood
$\mathcal U$ of the initial 
manifold $\mathbb R_x^d\times \{t=0\}$. In other words, if we consider for every
fixed $t\in\mathbb R$ the map
\begin{eqnarray}
\label{ray} \hat x(t,\cdot ) : \mathbb R^d&\rightarrow &\mathbb R^d\nonumber \\
x&\mapsto &\hat x(t,x) 
\end{eqnarray}
it is, in general, not one-to-one for large times $t\in\mathbb R$, see e.g. the
examples 1.2, 1.3 in section 5. 
Further it is known that the phase is discontinuous at \emph{caustics}, i.e.
points of intersection of rays, 
which in general will happen in finite times $t=t_{c_1}$, $t=t_{c_2}$, with
$t_{c_1}< 0< {t_{c_2}}$, 
called \emph{break-times}. \\
\emph{It is clear however that the formal WKB-expansion method can only be
justified for $t_{c_1}<t<t_{c_2}$} 
and thus a global, i.e. for all $t\in\mathbb R$, asymptotic description can not
be obtained from this method.\\

\noindent To overcome this deficiency, several generalizations of the method have been
developed in the past decades. 
Most of them rely on the use of \emph{global Fourier-Integral operators} (FIO) 
or the \emph{Maslov Canonical operator} (MCO),  which are now rather standard
methods and covered in many books, however for the sake of the reader which may be not 
familiar with it, we shall now give a flavour of these methods.\\


\subsection{The Fourier Integral Operator (FIO)}

\noindent Let us again start with an illustrative example: Consider the free Schr\"odinger equation in $\mathbb R^d$, i.e. 
\begin{align}
\varepsilon \partial _t\psi ^\varepsilon-i\frac{\varepsilon ^2}{2} \Delta \psi ^\varepsilon &=0,\\
\psi^\varepsilon  (t=0,x)&=\sqrt{n_I(x)}e^{iS_I(x)/\varepsilon }. 
\end{align}
Straightforward Fourier analysis shows, that its solution is explicitly given by the following oscillatory integral
\begin{equation}
\psi ^\varepsilon (x,t)=(2\pi \varepsilon )^{-d}\int_{\mathbb R}\int_{\mathbb R} \sqrt{n_I(z)}e^{iS(x,z,\xi ,t)/\varepsilon }dzd\xi,
\end{equation}
where
\begin{equation}
S (x,z,\xi ,t):=(x-z)\cdot \xi +\frac{t}{2}|\xi |^2+S_I(z).
\end{equation}
Let us recall the theorem of stationary phase (\cite{GuSt}, \cite{Mas1}, \cite{Mas2}, \cite{ReSi}):
\begin{proposition}
Let $a\in C_0^\infty (\mathbb R^d)$, $\Phi \in C^\infty(\mathbb R^d)$ and assume that the set $\{z: \nabla \Phi (z)=0,z\in \supp(a)\}$ 
consists of finitely many points $z_i$, with $i=1\dots N$. If the Hessians $D^2\Phi (z_i)$ are nonsingular, then for $\varepsilon \ll 1$
\begin{align*}
(2\pi \varepsilon )^{- d}\int_\mathbb {R^d} a(z)e^{i\Phi (z)/\varepsilon }dz =\nonumber \\
\sum _{i=1}^N \frac{1}{\sqrt{|\det D^2\Phi (z_i)|}}
\exp\left(\frac{i}{\varepsilon }\Phi (z_i)+\frac{i\pi }{4}m_i\right) (a(z_i)+O(\varepsilon )),
\end{align*}
where $z_i=z_i(x,t)$ and $m_i:=\sgn(D^2\Phi (z_i))$ is the so called \emph{Maslov index} of the the $i$-th ray.
\end{proposition}
\noindent This implies that (locally) the main contribution to the solution of the Schr\"odinger equation 
stems from stationary points w.r.t. $y$ and $\xi$, i.e. points at which $\nabla _{z,\xi }S=0$, which gives
\begin{equation}
\xi =\nabla S_I(z),\quad x=z+t\xi ,
\end{equation}
i.e. we get a ray-map, c.f. (\ref{ray}), defined by the following relation
\begin{equation}
\label{rel} x=\hat x(t,z)=z+t \nabla S_I(z).
\end{equation}
For small $t$ the map $z\mapsto \hat x(t,z)$ is singlevalued. In general however, there exist (maybe infinitely) many 
$z_i=z_i(x,t)$, which obey the equation (\ref{rel}). Note that the functions $S_I(z_i(x,t))$ are local solutions of 
the Hamilton Jacobi equation
\begin{equation}
\partial _tS +\frac{|\nabla _xS |^2}{2}=0.
\end{equation}
Provided that the Hessians $D^2S(z_i)$ are nonsingular 
and assuming that there exist only finitely many points $y_i$, the stationary phase theorem then gives us,
the following \emph{multivalued WKB-approximation} of $\psi ^\varepsilon $
\begin{equation*} 
\psi^\ep(x,t) \simeq  \sum_{i=1}^N \sqrt{\frac{n_I(z_i(x,t))}{|1+tD^2S_I(z_i(x,t))|}}
\exp\left(\frac{i}{\varepsilon }S_I(z_i(x,t) )+\frac{i\pi }{4}m_i\right) +O(\varepsilon ).
\end{equation*}
Note that the $m_i$ change sign, each time a ray passes the caustic, which is defined here to be a point at which 
$D^2S$ is singular.\\
\noindent For equations with variable coefficients, the above concepts have to be generalized, leading to the 
definition of a FIO acting on the initial datum $\psi^\eps_I(x)$ by 
\begin{equation*} \label{FIOpsi}
\psi^\varepsilon(x,t) \simeq 
(2\pi\varepsilon )^{-d} \int_{\R^d} \int_{\R^d} \A(x,z,\xi,t) e^{i \S(x,z,\xi,t)/\varepsilon } 
\psi^\ep_I(v) dz d\xi.
\end{equation*}
Since for $t=0$ the FIO  has to reduce to the identity operator, we obtain the conditions
\begin{equation}
\S(x,z,\xi,t)\vert_{t=0} = (x-z) \cdot \xi, \qquad 
\A(x,z,\xi,t)\vert_{t=0} = 1 \: .
\end{equation}
At the core of the use of FIOs is the fact that the composition and commutators
of two such FIOs is again "locally" a FIO (H\"ormander's Theorem \cite{Ho}) and 
that the rules to calculate their principial symbols are extensions of the 
rules of (composition and commutation of) pseudo differential operators.\\
The leading contribution again stems from the stationary points, which correspond to the asymptotic
solutions of the geometrical optics limit $\varepsilon\to 0$. 
For points on the caustic the FIO is an integral which can be brought into a canonical form and evaluated in 
terms of special integral functions. This is the problem of \emph{classification of Lagrangian singularities}
in contact geometry (see e.g. \cite{Ar}, \cite{AVG}). \\
\noindent For the detailed implementation of these basic ideas we refer 
e.g. to \cite{Du}, \cite{Fed}, \cite{GuSt}, \cite{Ho}, \cite{Kr}, \cite{La}, \cite{Lu}, \cite{Mas1}, \cite{Mas2}), 
however one should note that these approach generically assumes $S_I, n_I\in C_0^\infty(\mathbb R^d)$, which is of
course significantly stronger than our assumption (A2).\\ 

\noindent After reviewing these classical methods of high frequency approximation, we now turn to a new concept, namely the 
Wigner transformation technique.
\section{The Wigner function approach}

\begin{definition}
\noindent For given $f,g \in \mathcal S'(\mathbb{R}^d)$ and given $\varepsilon
\in (0, \varepsilon_0 ]$ 
we define the \emph{Wigner-transform} on the scale $\varepsilon $ by
\begin{equation} \label{Wigtrafo1}
w^\varepsilon (f,g)(x,\xi ) =
\frac{1}{(2\pi )^d}\int_{\mathbb{R}^d}f(x-\varepsilon \frac{z}{2})\bar
{g}(x+\varepsilon \frac{z}{2})e^{iz\xi }dz .
\end{equation}
\end{definition}
\noindent For fixed $\varepsilon$ the phase space-valued function
$w^\varepsilon$ is a continuous, 
bilinear mapping: 
\begin{equation*}
w^\varepsilon : \mathcal S'(\mathbb R_x^d)\times \mathcal S'(\mathbb
R_x^d)\longrightarrow 
\mathcal S'(\mathbb R_x^d\times \mathbb R_\xi^d )
\end{equation*}
In the following we shall use the notation: $w^\varepsilon [f]
=
w^\varepsilon (f,f)$.\\

\noindent The expression $w^\varepsilon [f ]$ is realvalued and for $f \in L^2 (\mathbb R_x^d)$ we 
conclude 
\begin{equation}
\| w^\varepsilon[f ] \| _2 = (4\pi  \varepsilon ^2)^{-d/2} \| \rho \| _2,
\end{equation}
where the \emph{density "matrix"} $\rho $ is defined by 
$\rho  (x,y):= f (x)\bar f (y)$.\\ 

\noindent\textbf{Remark.}  The Wigner-transform was originally introduced by 
E. Wigner in 1932 \cite{Wi} in the context of semi-classical Quantum
Mechanics.\\
We remark that there are slightly different definitions of the Wigner transform
in the literature depending essentially on the definition and normalization of
the 
underlying Fourier transform.\\

\noindent The inverse Wigner transform is given by 
\begin{equation}
\rho (x,y)\equiv f (x)\bar f (y)= (2\pi )^{-d} \int_{\mathbb R^d} w^\varepsilon [f]\left(\frac{x+y}{2}, \varepsilon \xi \right)e^{i\xi \cdot (x-y)} d\xi .
\end{equation}
Note that this transformation allows to obtain the function $f$ only up to a constant phase factor!\\

\noindent Now a simple calculation shows that formally
\begin{equation}
\label{zmom}\int_{\mathbb{R}^d }w^\varepsilon [f ](x,\xi,t) d\xi = |f(x,t)|^2,
\end{equation} 
which implies, that the energy density, as defined in (\ref{en}), is the zeroth moment of
$w^\varepsilon[\psi ^\varepsilon ]$ w.r.t. the velocity variable $\xi $. This is sometimes called
a \emph{microlocal decomposition of the energy} provided by theWigner transform. 
A rigorous justification (Note that $w^\varepsilon $ is in general not in $L^1$.) 
is given by (see \cite{LiPa} for details)
\begin{equation}
n^\varepsilon (x,t)=\lim_{\kappa \rightarrow 0}\int_{\mathbb R^d} w^\varepsilon [\psi ^\varepsilon ] (x,\xi ,t) e^{-\kappa |\xi |^2/2}d\xi ,
\end{equation}
which converges in $L^1$ towards $n^\varepsilon \in L^1_+(\mathbb R^d)$. 
More generally, an
important feature of the Wigner transform is that it facilitates a
"classical" computation of expectation values (mean values) of
physical observables $A^W(x,\varepsilon D)$ in any state $\psi ^\varepsilon$, namely  
\begin{equation}
\label{ccom}\left<\psi ^\varepsilon, A^W(x,\varepsilon D)\psi
^\varepsilon\right>_{L^2}=\int_{\mathbb{R}^d}\int_{\mathbb{R}^d}A(x,\xi
)w^\varepsilon [\psi^\varepsilon ](x,\xi)dxd\xi . 
\end{equation}
(Here we assume that the symbol $A(x,\xi )$ is in $\mathcal S(\mathbb R^{2d})$.)
In other words the real-valued function $w^\varepsilon $ acts as an equivalent
of the phase-space distribution function, however in contrast to classical phase
space distributions the Wigner transform is in general \emph{not point-wise
positive}! Indeed it has been proved for example in \cite{LiPa}, \cite{Hu} that
$w^\varepsilon[f ] \geq 0$ if and only if, either $f (x)=0$,
or $f (x)$ is a Gaussian.\\

\noindent \textbf{Remark.} It is well known (see e.g. \cite{LiPa}, \cite{MaMa}, \cite{Hu})
that averaging the Wigner function over phase space volumes large enough to
fulfill the \emph{Heisenberg uncertainty principle} yields a non-negative
function.
The so called \emph{Husimi functions} are obtained by convoluting the Wigner
function $w^\varepsilon [f]$ in $x$ and $\xi$ with the Gaussian 
\begin{equation*}
G^\varepsilon (z):=(\pi \varepsilon
)^{-\frac{d}{2}}\exp\left(-\frac{|z|^2}{\varepsilon }\right),
\end{equation*}
s.t. they become non-negative, i.e. $ w_H^\varepsilon (x,\xi ):=w^\varepsilon[f] \star_x G ^\varepsilon\star
_\xi  G^\varepsilon  \geq 0$ a.e..\\

\noindent For the sake of illustration we present some particular examples of
the "Wignerized" 
evolution equation (\ref{disp}):\\

\noindent\textbf{Examples:}\\

\noindent (i) The "Wignerized" Schr\"odinger equation
(\ref{schro})  or \emph{Wigner equation} (for details on this equation see e.g.
\cite{GaMa}, \cite{MaMa})
\begin{equation} \label{Wigeq}
\partial _tw^\varepsilon+\xi \cdot \nabla_x w^\varepsilon -\theta ^\varepsilon
[V]w^\varepsilon =0,\quad x, \xi \in \mathbb R^d, t\in\mathbb R
\end{equation}
where $\theta^\varepsilon  [V]$ is the (non-local) pseudo-differential operator
\begin{align}
\theta^\varepsilon  [V]w^\varepsilon(x,\xi ,t) := \frac{i}{(2\pi )^d\varepsilon
}\int_{\mathbb R^d}\int_{\mathbb R^d}[V(x+\varepsilon
\frac{z}{2})-V(x-\varepsilon \frac{z}{2})] \nonumber
\\ \cdot  w^\varepsilon(x,\xi ,t)\ e^{iz (\xi -\xi')}dzd\xi ' ,
\end{align}
and $w^\varepsilon (t):=w^\varepsilon [\psi^\varepsilon (t)]$, where $\psi^\varepsilon $ solves 
(\ref{disp}), (\ref{ho}). 
Note that in the free motion case, i.e. $V(x)\equiv 0$, the Wigner equation becomes the free transport
equation of classical statistical mechanics
\begin{equation*}
\partial _tw^\varepsilon+\xi \cdot \nabla_x w^\varepsilon =0,\quad x, \xi \in
\mathbb R^d; t\in\mathbb R.
\end{equation*}
Further note that in the case of potentials $V$ which are quadratic in $x$ the
operator $\theta^\varepsilon  [V]$ simplifies to the classical force term
$\nabla _x V$ for all positive $\varepsilon $. The harmonic oscillator,
$V(x)=|x|^2/2$, is a typical example
\begin{equation*}
\partial _tw^\varepsilon+\xi \cdot \nabla_x w^\varepsilon -x\cdot \nabla _\xi
w^\varepsilon =0,\quad x, \xi \in \mathbb R^d, t\in\mathbb R.
\end{equation*}
\noindent (ii) The "Wignerized" 1-d Airy equation (\ref{airy})
\begin{equation}
\partial _tw^\varepsilon +\xi ^2\partial _xw^\varepsilon-\frac{\varepsilon
^2}{12}w _{xxx}^\varepsilon =0,\quad x,\xi , t\in\mathbb R.
\end{equation}
This equation is obtained after a straightforward, but lengthy calculation.
\hfill$\diamondsuit $\\

\noindent We shall now perform the $\varepsilon \rightarrow 0$ limit associated to the
dispersive equation (\ref{disp}) with initial data (\ref{ho}), assuming (A1),
(A2).\\ 


\subsection{The WKB-limit using Wigner transforms} 

\noindent One of the crucial properties (for a proof see \cite{GMMP}) of the
Wigner 
transform is 
\begin{proposition}\emph{\cite{GMMP}}
If $f$ and $g$ lie in a bounded subset $\mathcal B$ of $L^2(\mathbb R^d)$, then
$w^\varepsilon (f,g)$ is bounded uniformly in $\mathcal S'(\mathbb R_x^d\times
\mathbb R_\xi ^d)$ as $\varepsilon \rightarrow 0$. More precisely we have 
\begin{equation}
(\mathcal F_{\xi \rightarrow z}w^\varepsilon (f,g))(x,z)\in C_0(R^d_z;
L^1(\mathbb R^d))
\end{equation}
uniformly for $f,g\in\mathcal B$ as $\varepsilon \rightarrow 0$.
\end{proposition}
\noindent Thus, by compactness, there exists a sub-sequence $\varepsilon _k$ and
a distribution $w^0\in\mathcal S'(\mathbb R^d_x\times \mathbb R^d_\xi )$ such
that
\begin{equation}
w^{\varepsilon _k}[f^{\varepsilon _k}]\stackrel{k\rightarrow \infty
}{\longrightarrow}w^0 \mbox{ in $\mathcal S'(\mathbb R_x^d\times \mathbb R_\xi
^d)$.}
\end{equation}
\noindent It has been shown (for example in \cite{LiPa},\cite{MaMa}) that the
limiting points of the Husimi function $w_H^\varepsilon $ are also the limiting points of the 
Wigner function. Since $w^\varepsilon _H$ is non-negative, this implies that also $w^0$ 
\emph{is non-negative}. More precisely it is a positive Borel-measure, i.e. $w^0\in \mathcal M^+$, 
(the cone of bounded positive Borel measures) 
and therefore can be interpreted indeed as a classical phase-space measure,
called the \emph{Wigner measure} of $f^{\varepsilon _k}$ (See \cite{Ge}, \cite{GMMP} for a
proof of the positivity of $w^0$.). \\

\noindent \textbf{Remark.} The \emph{Wigner measures} are a particular version of $L^2$
defect measures, related to the \emph{H measures} of
L. Tartar \cite{Ta} and P. G\'erard \cite{Ge}. For more details see the 
expository article of N. Burq \cite{Bu}.\\ 

\noindent For simplicity we now restrict the scale
$\varepsilon $ to a sub-sequence, such that the Wigner measure $w^0$ is unique, 
i.e. independent of the choice of the subsequence, and in the following we denote it by $w=w^0$.\\
We further introduce  the $d$-dimensional \emph{Poisson-bracket} of two functions $f,g$, i.e.
\begin{equation}
\{f,g\}:= \nabla _x f \cdot \nabla _\xi g -\nabla _\xi  f \cdot \nabla _x g  .
\end{equation}
\noindent In  proposition 3.2 below we state the well known fact that the Wigner
transform translates the action of an Weyl-operator asymptotically in zeroth
order into a multiplication and in first order into a Poisson bracket. (A proof
can be found in the appendix of \cite{GMMP})
\begin{proposition}\emph{\cite{GMMP}}
Let $P \in S^\sigma $ for some $\sigma \geq 0$. Then if $f,g$ lie in a bounded
subset $\mathcal B$ of $L^2(\mathbb{R}^d)$, the expansion
\begin{equation}
\label{exp}w^\varepsilon (P^W(x,\varepsilon D)f,g)=Pw^\varepsilon (f,g)+\frac{\varepsilon
}{2i}\{P,w^\varepsilon (f,g)\}+O(\varepsilon ^2)
\end{equation}
holds in $\mathcal S'(\mathbb{R}^d_x\times \mathbb{R}^d_\xi )$ uniformly for
$f,g\in\mathcal B$.
\end{proposition}
\noindent Now let $\psi^\varepsilon $ satisfy the IVP (\ref{disp}), (\ref{ho}). 
In view of  (\ref{cons}) and proposition 3.1 one obtains the uniform
boundness of  $w^\varepsilon [\psi ^\varepsilon ] (t)$ in $L^\infty (\mathbb{R};
\mathcal S'(\mathbb R_x^d\times \mathbb R_\xi^d ))$ and thus the existence of
$w\in L^\infty (\mathbb{R};\mathcal M^+(\mathbb R_x^d\times \mathbb R_\xi^d ))$
such that  
\begin{equation}
w^\varepsilon[\psi^\varepsilon ]\stackrel{\varepsilon \rightarrow
0}{\longrightarrow}w(dx,d\xi, t)\mbox{ in $L^\infty (\mathbb{R};\mathcal
M^+(\mathbb R_x^d\times \mathbb R_\xi ^d))$ weak-$\star $.} 
\end{equation}
\noindent\textbf{Remark.} It has been shown in \cite{Ge}, \cite{GMMP},
\cite{LiPa} that the limiting process is actually locally uniform in t, i.e.
$w\in C_b(\mathbb{R}; \mathcal M^+ (\mathbb R_x^d\times \mathbb R_\xi^d
)w-\star)$. \\ 

\noindent To derive an evolution equation for the limiting phase space
distribution $w$ 
we differentiate 
\begin{equation*}
\partial _t w^\varepsilon =
w^\varepsilon (\psi ^\varepsilon _t,\psi ^\varepsilon )+w^\varepsilon (\psi
^\varepsilon ,\psi_t ^\varepsilon )=
2\mbox{ Re }w^\varepsilon (\psi _t^\varepsilon ,\psi ^\varepsilon ) .
\end{equation*}
Now using equation (\ref{disp}) and proposition 3.2, having in mind that
$w^\varepsilon $ and 
$H(x,\xi )$ are real-valued, we obtain the following essential equation
\begin{equation}
\partial _t w^\varepsilon =
-\{H,w^\varepsilon\}+O(\varepsilon),
\end{equation}
which is a linear PDE in $w^\varepsilon $. 
Passing to the limit $\varepsilon \rightarrow 0$ yields, in the sense of
distributions, 
the classical \emph{Liouville equation}
\begin{equation} \label{liou}
\partial _t w+\{H,w\}=0,\quad x\in \mathbb R^d, t\in\mathbb R.
\end{equation}
We calculate the corresponding initial data $w_I$ which is the limit of the
Wigner transform 
of the WKB-initial data $\psi ^\varepsilon _I:=\sqrt{n _I}e^{i S
_I/\varepsilon}$ 
\begin{eqnarray} \label{transwkb}
w^\varepsilon[\psi^\varepsilon _I](x,\xi )=
\frac{1}{(2\pi )^d}\int_{\mathbb R_z^d}\sqrt{n_I(x-\varepsilon
\frac{z}{2})n_I(x+\varepsilon \frac{z}{2})}\cdot \nonumber\\
\mbox{exp}\left[\frac{i}{\varepsilon }\left(S_I(x-\varepsilon
\frac{z}{2})-S_I(x+\varepsilon \frac{z}{2})\right)+iz\xi \right]dz .
\end{eqnarray}
\begin{lemma} Let $w^\varepsilon[\psi^\varepsilon _I]$ be given by
(\ref{transwkb}), then
\begin{equation}\label{mono}
w^\varepsilon[\psi ^\varepsilon_I]\stackrel{\varepsilon \rightarrow 0
}{\longrightarrow} w_I:=
n_I(x) \delta (\xi -\nabla_x S_I(x))dx.
\end{equation}
\end{lemma}
\begin{proof} The proof is a straightforward computation having in mind that the
amplitude $n_I(x)$ and the phase $S_I(x)$ are by assumption
$\varepsilon$-independent.\\
\end{proof}
\noindent\textbf{Remarks.}\begin{itemize} 
\item This special type of initial phase space distributions (\ref{mono}) are
called 
\emph{mono-kinetic initial data}, i.e. for every $x\in \mathbb R^d$ there is
exactly 
one (characteristic) speed $v_I(x):=\nabla S_I(x)$. 
Note that we have chosen the scale of oscillations in the initial data to be
\emph{equal} 
to the small parameter $\varepsilon $ in the equation (\ref{disp}). \\ 
We further remark that in case $S_I(x)\in C^\infty $ the set $\{(x,\nabla
_xS_I(x))\}$ 
is a \emph{Lagrangian submanifold} of phase-space $\mathbb R^{2d}$, 
i.e. a manifold on which the symplectic form $\omega :=dx\wedge d\xi $ vanishes 
(for details see e.g. \cite{Fed}, \cite{Ho}).
\item The Wigner measure approach works for much more general initial data (cf
\cite{GMMP}, 
in this work we restrict to WKB data for the sake of comparison.
\end{itemize}

\noindent Further we have:
\begin{lemma}\emph{\cite{GMMP}} Let $w^\varepsilon[\psi(t)]$ be the Wigner
transform of the solution of  
(\ref{disp}) (\ref{ho}) (calculated from \fer{liou}, \fer{mono}) 
and assume (A1), (A2), then the associated density $n^\varepsilon (x,t)$ given
by (\ref{zmom}) satisfies
\begin{equation}
n^\varepsilon (x,t)\stackrel{\varepsilon \rightarrow 0 }{\longrightarrow} 
n(x,t):=\int_{\mathbb R^d}w(x,d\xi ,t)\in C_b(\mathbb R^d; \mathcal M^+(\mathbb
R^d))
\end{equation}
where the convergence is locally uniform in $t$.
\end{lemma}
\begin{proof} A proof can be found for example in \cite{GaMa}, \cite{GMMP},
\cite{LiPa}. Note that the technical assumptions of \emph{"$\varepsilon
$-oscillatory"} and \emph{"compact at infinity"} initial data,
which are introduced in \cite{Ge}, \cite{GMMP}, are fulfilled if one imposes
(A2).
\end{proof}
\noindent \textbf{Remark.} One can show that the $\varepsilon \rightarrow 0$
limit of other observables $A^W$, or more general expressions which are
quadratic in $\psi ^\varepsilon $ with a specific growth in $\xi $, can be
computed essentially in the same way, i.e. the limit is given by the right hand
side of (\ref{ccom}) with $w^\varepsilon $ replaced by $w$. For details see
again \cite{GaMa}, \cite{GMMP}, \cite{LiPa}. \\

\noindent Thus by using the Wigner transform we obtained a (semi-) classical
phase-space description, which we shall analyse in the next section.


\section{Analysis of the Wigner measure}

This is the main part of the work, in which we will try to establish the precise
relation between WKB-asymptotic solutions of (\ref{disp}) and the Wigner measure
that has been obtained in the last section. We start with the following
definition:
\begin{definition}
The Hamiltonian flow $F_t$ associated to the Liouville equation (\ref{liou}) is given by
\begin{equation}
F_t(x,\xi )= (\tilde x(t,x,\xi),\tilde \xi (t,x,\xi))
\end{equation}
where $(\tilde x,\tilde \xi )$ solve the initial value problem
\begin{eqnarray}
\frac{d{\tilde x}}{dt} &=& \nabla _{\xi} H({\tilde x},{\tilde \xi }),
\quad \tilde{x}(0,x,\xi )= x\\
\frac{d {\tilde \xi}}{dt}  &=& -\nabla _{y} H({\tilde x},{\tilde \xi
}),\quad \tilde{\xi }(0,x,\xi )= \xi. 
\end{eqnarray}
\end{definition}
\noindent \textbf{Remark.} The above ODE's are usually referred to as
\emph{Hamilton's equations} and the curves $(\tilde x,\tilde \xi )$ are often
called \emph{bicharacteristics}. \\

\noindent For the sake of simplicity we shall assume in the sequel:\\

\noindent\textbf{Assumption (A3)}\\

\noindent \emph{The Hamiltonian flow $F_t$ is a continuously differentiable
globally defined map for every $t\in\mathbb R$.\\}

\noindent\textbf{Remark.} The global existence of $F_t$ is a priori not
guaranteed for general $H\in S^\sigma$, 
except for $\sigma \leq 1$. 
(For more details see for example \cite{Ar}, \cite{ReSi}.) 
We further remark that well known situations for which (A3) is valid, 
like the case of $\alpha $-elliptic Hamiltonians $H\geq C_1|\xi |^\alpha-C_2$
with $C_1,C_2,\alpha >0$ 
are covered by our theory. 

\noindent An illustrative example where the Hamiltonian flow is not global in
time is given in Appendix 1.\\

\noindent A straightforward calculation shows that the \emph{Hamiltonian
function} $H(x,\xi )$ is constant along the  flow $F_t$, i.e.
\begin{equation}
\label {hcons}H(\tilde x(t,x,\xi ),\tilde \xi(t,x,\xi ) )=H(x,\xi ),\quad\forall
\ t\in\mathbb R, x,\xi \in\mathbb R^d.
\end{equation} 
\noindent Also, by the classical \emph{Theorem of Liouville} (see e.g.
\cite{Ar}), we have that $F_t$ is \emph{volume preserving}; 
i.e. its Jacobian satisfies 
\begin{equation}
\det\left(\frac{\partial (\tilde x(t,x,\xi),\tilde \xi (t,x,\xi))}{\partial
(x,\xi)}\right)=1,\quad\forall \ t\in\mathbb R, x,\xi \in \mathbb R^d.
\end{equation}
With the above definitions the method of characteristics guarantees that the
unique (weak) solution of the Liouville equation (\ref{liou}) subject to the
initial condition $w(t=0)=w_I\in\mathcal M^+(\mathbb R^d_x\times \mathbb R^d_\xi
)$ satisfies 
\begin{equation}
\label{ide}\int_{\mathbb{R}^d}\int_{\mathbb{R}^d}\varphi (x,\xi ) w(dx,d\xi,
t)=\int_{\mathbb{R}^d}\int_{\mathbb{R}^d}\varphi (F_t(x,\xi )) \ w_I(dx,d\xi )
\end{equation}
for all test functions $\varphi \in C_b(\mathbb{R}_x^d\times \mathbb{R}_\xi ^d)
$. In other words the solution $w$ of (\ref{liou}) remains constant along all
bicharacteristic curves $(\tilde x,\tilde \xi)$. Since (\ref{liou}) is a
\emph{linear} equation and (A3) is assumed we know that the above solution $w$
exists for all $t\in\mathbb R$. In contrast to the case of a nonlinear first
order PDE we have \emph{global} solutions for equation (\ref{liou}); i.e.
\emph{no caustics appear in phase space}! This is sometimes referred to as
\emph{"unfolding" of caustics.}\\

\noindent In the next subsection we connect the WKB-system (\ref{wkb1}),
(\ref{wkb2}) with a special class of solutions to the Liouville-equation, which we shall link to 
the traditional WKB method in the pre-caustic regime.


\subsection{On mono-kinetic phase space distributions}

\noindent Taking into account the monokinetic form of the initial data, we
observe that, with the use of equation (\ref{mono}), identity (\ref{ide}) reads
\begin{equation}
\label{ide2}\int_{\mathbb R^d}\int_{\mathbb R^d}\varphi (x,\xi ) \ w(dx,d\xi
,t)=\int_{\mathbb R^d}n_I(x)\varphi (\hat x(t,x),\hat \xi (t,x)) \ dx,
\end{equation} 
for all $\varphi \in C_b(\mathbb{R}^d_x\times \mathbb{R}_\xi ^d)$ where the
curves $(\hat x(t,x) ,\hat \xi (t,x))$ solve the IVP (\ref{hode1}),
(\ref{hode2}). Note that, since the initial data are given by $\xi=\nabla
_xS_I(x)$, the curves $\hat x(t,x)$ are the rays associated to the
Hamilton-Jacobi equation 
\begin{equation*}
\partial  _t S + H(x,\nabla _xS )=0,\quad S (x,0)=S_I(x).
\end{equation*}
\noindent The above reflects the fact that the $(\tilde x,\tilde \xi )(t,
x,\xi)$ bicharacteristics in phase space are projected down to rays $\hat
x(t,x)$ in position space plus an additional curve $\hat \xi (t,x)$, which is
the gradient of the phase $S$ along the rays. More precisely, we have

\begin{equation}
(\hat x,\hat \xi )(t,x)=(\tilde x,\tilde \xi )(t,x,\nabla _xS_I(x)).
\end{equation}
The function $S$ is often called the \emph{generating function of the flow} $F_t$.\\

\noindent We shall use the following lemma for the proof of theorem 4.1.
\begin{lemma} Assume (A1)-(A3) and let $w(t)$ be a (weak) solution of
(\ref{liou}) with mono-kinetic initial data $w_I$ (\ref{mono}). Then $\forall$
$t\in \mathbb{R}$
\begin{equation}
\label {lem}\int_{\substack{\mathbb{R}^d}}\int_{\substack{\mathbb{R}^d}} \phi
(H(x,\xi ))  \ w(dx,d\xi, t)
=\int_{\substack{\mathbb{R}^d}} \phi (H(x,v_I(x))) \ n_I(dx)
\end{equation}
for every real valued continuous and bounded from below function $\phi$ for
which the right hand side of (\ref{lem}) is finite. 
\end{lemma}
\begin{proof} At first take $\phi \in C(\mathbb R)$ bounded. Setting $\varphi
(x,\xi )=\phi (H(x,\xi ))$ in (\ref{ide2}) combined with the fact that $H(\tilde
x,\tilde \xi )$ is conserved proves the claim. For unbounded $\phi $ a density
argument gives the result.
\end{proof}
\noindent\textbf{Remark.} If the Hamiltonian is bounded from below and gauged 
such that $H(x,\xi )\geq 0$ the above lemma states the energy conservation
property of 
the Liouville equation, when setting $\phi (\cdot )= \id$ on $\mathbb R^+$.\\

\begin{definition} We define the following set of functions: 
\begin{equation} 
\Lambda :=\{\lambda\in C(\mathbb R;\mathbb R^+): \int_{\mathbb R^d}\int_{\mathbb R^d}\lambda (H(x,\xi )) \ w_I(dx,d\xi)<\infty\}.
\end{equation}
\end{definition}

\noindent Now the following theorem holds:
\begin{theorem}Assume (A1)-(A3) and let $-\infty \leq t_{c_1}<0<
t_{c_2}\leq \infty$.\\
\emph{(i)} The unique (weak) solution of $(\ref{liou})$ 
\begin{equation*}
\partial _t w+\{H,w\}=0\quad\mbox{in $\mathcal D'(\mathbb R_x^d\times \mathbb
R^d_\xi \times(t_{c_1},t_{c_2})) $}
\end{equation*}
is given by 
\begin{equation}
\label{monosol}w(x,\xi ,t)=n(x,t)\delta(\xi-v(x,t)),
\end {equation}
if and only if the pair $(n,v)(x,t)$, where $n\in C_b((t_{c_1},t_{c_2});
\mathcal M^+(\mathbb{R}^d))$ and $v\in L^\infty
((t_{c_1},t_{c_2});L^1(\mathbb{R}^d;dn(t))^d)$ is a solution of
\begin{equation}
\label{renorm}\partial _t(n\sigma (v))+\diverg(n\sigma (v)\nabla _\xi
H(x,v))+n\nabla _\xi \sigma (v)\nabla _yH(x,v)=0
\end{equation}
in $\mathcal D'(\mathbb{R}^d\times(t_{c_1},t_{c_2}))$, for every $\sigma \in
C^1(\mathbb{R}^d )$ for which there exists a $\lambda \in\Lambda $ such that:
\begin{equation}
\label {scon}\frac{\vert \sigma \vert}{1+\lambda (H)}\in L^\infty (\mathbb R^d),
\frac{\vert \sigma \vert \vert \nabla _\xi H \vert }{1+ \lambda (H) }\in
L^\infty (\mathbb R^d),  \frac{\vert \nabla _\xi \sigma \vert \vert \nabla _y H
\vert}{1+ \lambda (H) }\in L^\infty (\mathbb R^d).
\end{equation}
\noindent \emph{(ii)} Moreover if $H$ is such that $\sigma \equiv 1$ and $\sigma
\equiv v_i$, $i=1\dots d$ satisfy (\ref{scon}), then the monokinetic solution
(\ref{monosol}) is uniquely determined by the following system of fluid-type
equations
\begin{eqnarray}
\label{euler1}\partial _tn+\diverg(n\nabla _\xi H(x,v))&=&0, \\
\label{euler2}\partial _t(nv)+\diverg(\nabla _\xi H(x,v)\otimes nv)+n\nabla
_yH(x,v)&=&0
\end{eqnarray}
in $\mathcal D'(\mathbb R^d\times(t_{c_1},t_{c_2}))$ with initial data  
\begin{equation}
n(x,0)=n_I(x), \quad v(x,0)=v_I(x).
\end{equation}
\end{theorem}
\noindent The equation (\ref{renorm}) is called a \emph{generalized
formulation} of the system (\ref{euler1}), (\ref{euler2}). It reflects the fact, that different 
moments of the monokinetic Wigner measure can be chosen, which is sometimes useful 
for numerical purposes.
\begin{proof}
Choose a test function $\varphi =\gamma (x,t)\sigma (\xi )\in \mathcal D(\mathbb
R_x^d\times \mathbb R_\xi ^d \times (t_{c_1},t_{c_2}))$, then the weak
formulation of (\ref{liou}) reads:
\begin{eqnarray*}
\int_{t_{c_1}}^{t_{c_2}} \int_{\mathbb R^d}\int_{\mathbb R^d}[( \sigma (\xi
)\partial _t+\sigma (\xi )\nabla _\xi H(x,\xi )\cdot \nabla _x)\gamma(x,t)\\ 
-\nabla _yH(x,\xi )\cdot \nabla _\xi \sigma (\xi )\gamma(x,t)]\ w(dx,d\xi,dt)=0
\end {eqnarray*}
by inserting $w$ given by (\ref{monosol}) we obtain:
\begin{eqnarray*}
\int_{t_{c_1}}^{t_{c_2}}\int_{\mathbb R^d}  [(\sigma (v(x,t))\partial _t+\sigma
(v(x,t))\nabla _\xi H(x,v(x,t))\cdot \nabla _x)\gamma(x,t)\\ 
-\nabla _yH(x,v(x,t))\cdot \nabla _\xi \sigma (v(x,t))\gamma(x,t)] \ n(dx,dt)=0
\end {eqnarray*}
which is exactly the weak formulation of (\ref{renorm}).
Now let $\sigma _l\in \mathcal D(\mathbb{R}^d_\xi )$ be a sequence of test
functions satisfying the conditions stated in the theorem and converging a.e.
to $\sigma \in C^1(\mathbb{R}^d_\xi )$ as $l\rightarrow \infty $. Because the
test functions $\sigma_l $
satisfy the conditions above (\ref{scon}), the dominated convergence theorem
(with the use of lemma 4.1. and definition 4.2. implies the assertion (ii). Now
we can choose successively $\sigma \equiv 1$ and $\sigma \equiv v_i$, $i=1,...,
d$ in equation (\ref{renorm}) to obtain the system (\ref{euler1}),
(\ref{euler2}) .\\
\end{proof}
\noindent \textbf{Remark.} Note that in case $H$ is given by 
\begin{equation}
\label{sph}H(x,\xi )=\frac{|\xi |^2}{2}+V(x),\quad x,\xi \in\mathbb R^d,
\end{equation}
where $V$ is bounded from below we can choose $\lambda(\cdot ) = \id$ and the above
system of fluid-type equations (\ref{euler1}), (\ref{euler2}) simplifies to the
well known \emph{zero-temperature Euler-equations} of rarefied gas dynamics
\begin{eqnarray}
\partial _tn+\diverg(nv)&=&0,\quad x\in\mathbb R^d,t\in \mathbb
(t_{c_1},t_{c_2}) \label{Eul1}\\
\partial _t(nv)+\diverg(nv\otimes v)+n\nabla _xV&=&0 \label{Eul2}, \\
n(x,0)=n_I,\quad v(x,0)=v_I,
\end{eqnarray}
which is, for $t_{c_1}<t<t_{c_2}$, the $\varepsilon \rightarrow 0$ limit 
of the QHD system (\ref{qhd1}), (\ref{qhd2}) that has been mentioned in the introduction.\\

\noindent We now establish the precise connection between the WKB-system
(\ref{wkb1}), (\ref{wkb2})
and the above theorem in the following corollary.
\begin{corollary} Let  $(t_{c_1}, t_{c_2})$  be as above;\\
(i) Further let $S\in C^2(\mathbb R^d \times (t_{c_1}, t_{c_2}))$ be a smooth
solution of the  Hamilton-Jacobi equation (\ref{wkb2})
\begin{equation*}
\partial _tS+H(x, \nabla _xS )=0,\quad S(x,0)=S_I(x).
\end{equation*}
Define $v(x,t):=\nabla _xS(x,t) $, and let $n$ be the unique solution of
(\ref{euler2}), then $(n,v)$ satisfies (\ref{euler1}), (\ref{euler2}) and is a
generalized solution in the sense defined above.\\
(ii) \noindent Let $(n,v)\in C^1(\mathbb R^d \times (t_{c_1}, t_{c_2}))$ be a
smooth solution of (\ref{euler1}), (\ref{euler2}).\\
If the initial velocity is given by $v_I(x)=\nabla _xS_I(x)\in C^1(\mathbb
R^d)$, then there exists a phase function $S(x,t)$, unique up to a constant,
which is a solution of (\ref{wkb2}) on the same time interval. In particular the velocity $v$ is a
gradient field and the solution of (\ref{liou}) can be written in the form
\begin{equation*}
\label{monosol2} w(x,\xi ,t)=n(x,t)\delta(\xi-\nabla _xS(x,t))\in\mathcal
C_b((t_{c_1}, t_{c_2});M^+(\mathbb R_x^d\times \mathbb R_\xi ^d).  
\end {equation*}
\end{corollary}
\begin{proof} Differentiating (\ref{wkb2}) w.r.t. $x_i$ and using the chain rule (which is
rigorous because we are in the regime of classical solutions) yields 
\begin{equation*}
\partial ^2_{t x_i}S+\frac {\partial }{\partial y_i}H(x,\nabla _xS )+\nabla_\xi
H(x,\nabla _xS )\cdot \frac {\partial }{\partial x_i}\nabla _x S   =0
\end{equation*}
and, with $v_i:=\frac {\partial }{\partial x_i}S $ we get 
\begin{equation}
\label{genb}\partial _tv+(\nabla_\xi H(x,v)\cdot \nabla _x) v+\nabla _y H(x,v )   =0.
\end{equation}
We multiply by $\nabla _v\sigma (v)$ to obtain
\begin{equation*}
\sigma (v) _t+(\nabla _\xi H(x,v)\cdot \nabla _x)\sigma (v)+\nabla _v \sigma (v)
\nabla _yH(x,v )  =0.
\end{equation*}
Again multiplying this equation by a regular solution $n$ of the continuity
equation (\ref{euler1}) yields (\ref{renorm}). Hence choosing appropriate test
functions we have proved that $(n,v)$ is a generalized solution on
$(t_{c_1},t_{c_2})$. We proove claim (ii), by first eliminating $n$ from equation (4.12). 
Using (4.11) we obtain on
$(t_{c_1}, t_{c_2}) $ the equation (\ref{genb}) above subject to 
$v(x,0)=\nabla _xS_I(x)$. 
One checks that the characteristic ODE system for this PDE is given by
(\ref{hode1}), (\ref{hode2}). 
Thus we can identify $v(\hat x(t,x),t)=\hat \xi (x,t)$ and conclude the existence 
of a unique (up to a constant) function $S(x,t)$ s.t. $v=\nabla _xS$. 
The rest of the proof is identical to the one of (i) read from bottom up.
\end{proof}
\noindent Note that \emph{the system (\ref{euler1}),
(\ref{euler2}) 
is defined $n(t)$-a.e. and thus remains valid as long as the density is
single-valued} ! 
In particular this holds in the examples 1.2 , 2 in section 5 below.


\subsection{Multiple phases}

\noindent We will now show, under some more stringent assumptions, that
\emph{away from the caustic} we can \emph{locally} extend $w(t)$ beyond
$t_{c_{1,2}}$ as a sum over mono-kinetic distributions, which will lead to a
generalization of the asymptotic expression of the solution $\psi^\varepsilon
(t)$ of (\ref{disp}).\\ 

\noindent Since  the Wigner measure remains constant along the flow $F_t$, we
can write for $\varphi \in C_b(\mathbb R_x^d\times \mathbb R_\xi ^d)$ 
\begin{equation}
\left<w(t),\varphi \right>=\left<w_I(F_{-t}), \varphi
\right>=\left<n_I(\tilde x(-t,x,\xi ))\delta (f_{x,t}(\xi )), \varphi 
\right>
\end{equation}
where $\left<\cdot ,\cdot \right>$ denotes the duality bracket in the
sense of measures and further we have used the following definition:
\begin{definition}
The mapping $f_{x,t}(\cdot )\in C^1: \mathbb R^d \rightarrow \mathbb R^d$ is
defined by
\begin{equation}
f_{x,t}(\xi ):=\tilde \xi(-t,x,\xi )-\nabla _xS_I(\tilde x(-t,x,\xi )). 
\end{equation}
\end{definition}
\noindent Above $(x,t)$ are parameters as the notation indicates.

\begin{definition} We denote the nullset of $f_{x,t}(\cdot )$ by
\begin{equation}
\label{kern}\mathcal K(x,t):=\{\xi \in\mathbb R^d: f_{x,t}(\xi )=0\}
\end{equation}
and the corresponding functional determinant by
\begin{equation}
Df_{x,t}(\xi ):=\det \left(\frac{\partial f_{x,t}(\xi )}{\partial \xi }\right).
\end{equation}
\end{definition}
\noindent In general we cannot hope for $f_{x,t}(\cdot )$ to be a diffeomorphism
in the whole propagation domain, at least we can get local results if we
impose:\\

\noindent \textbf{Assumption (A4)}\\

\noindent \emph{Let $\mathcal U\subseteq \mathbb R^d_x\times \mathbb R_t$ be such an
open set 
and $N\in\mathbb N$ be such an integer that the nullset (\ref{kern}) can be
written as}
\begin{equation} \label{Krep}
\mathcal K(x,t)=\bigcup_{i=1}^{N} \{v_i(x,t) \}\quad\mbox{\emph{for all } $(x,t)\in
\mathcal U$.}
\end{equation}

\noindent This means that we only
allow situations with a \emph{finite} number $N$ of (gradients of) phases at
each point of the propagation domain. 
Of course $w$ is well defined even if (A4) does not hold, however in general no
upper bound $N$ can be found, $N=\infty $ is possible!\\

\noindent \textbf{Remark.} 
Indeed very little is known about this problem in general situations, so far the
only well studied example is the one dimensional  free motion case, i.e.
$H=H(\xi )$, $\xi \in \mathbb R$. Under the additional assumption that $H(\xi )$
is strictly convex, i.e. there exists $c>0$ such that $H^{''}(\xi )\geq c$,
Tadmor and Tassa have shown in \cite{TaTa}, that if $H'(S'_I(x))$ has a
\emph{finite number of inflection points}, then the number number of
"original" shocks, i.e. shocks that do not result from the interaction
of other shocks, in the entropy solution to the corresponding conservation law
is finite, which implies (A4). 
An example of an initial condition $\nabla S_I$ which evolves, in the particular
case of $H=\xi^2/2$, into an a.e. $C^\infty $ function with \emph{countably many
original shocks} can be found in \cite{Sch}.
\begin{theorem} \label{theorem4.2}
Assume (A1)-(A4) and denote by $v_i(x,t)$,  $i=1\dots N$, the elements of the
nullset $\mathcal K(x,t)$.\\
If the point $(x,t)\in\mathcal U$ is such, that $Df_{x,t}(v_i)\not=0$ for all
$i=1\dots N$, 
then the measure-valued solution of the Liouville equation (\ref{liou}) can be
written as
\begin{equation}\label{multiphas}
w(x,\xi ,t) = \sum_{i=1}^{N}\frac{n_I\left(\tilde x(-t,x,\nabla
_xS_i(x,t))\right)}
{|Df_{x,t}(\nabla _xS_i(x,t))|}\ \delta (\xi -\nabla _xS _i(x,t))
\end{equation}
where $S_i\in C_{loc}^2(\mathcal U)$ such that $\nabla _xS_i=v_i\in\mathcal
K(x,t)$ for all $i\in \{1\dots N\}$.
\end{theorem}
\begin{proof} Choose tests functions $\varphi (x,\xi ,t)=\gamma (x,t) \sigma
(\xi )\in C_b(\mathbb R_t;C_b(\mathbb R^{2d}))$. 
As above we write
\begin{equation}
\label{dide}\left<w(t),\varphi \right>
=\left<n_I(\tilde x(-t,x,\xi ))\delta (f_{x,t}(\xi )), \varphi  \right>.
\end{equation}
\noindent By the coarea formula (see \cite{Ev}, \cite{Fe}) we have 
\begin{equation*}
\left<\delta \left(f_{x,t}(\xi )\right),\sigma (\xi ) \right>
=\left<\delta (\zeta ),\bar \sigma_{x,t}(\zeta ) \right>
\end{equation*}
with
\begin{equation}
\bar \sigma_{x,t}(\zeta ):=\int_{\{f^{-1}_{x,t}(\zeta )\}}\sigma (\xi
)\frac{d\xi }{|Df_{x,t}(\xi )|},
\end{equation}
which is well defined as long as $Df_{x,t}(\xi )\not=0$, even if $f_{x,t}(\xi )$
is not an isomorphism for all points $\xi \in\mathbb R^d$! By the definition of
the delta distribution the above is equal to
\begin{equation*}
\left<\delta (\zeta ),\bar \sigma(\zeta )\right>\equiv \bar \sigma
(0)=\int_{\mathcal K(x,t)}\sigma (\xi )\frac{d\xi }{|Df_{x,t}(\xi )|},
\end{equation*}
since $\{f^{-1}_{x,t}(0)\}=\{\xi \in\mathbb R^d: f_{x,t}(\xi )=0 \}=:\mathcal
K(x,t)$. By assumption (A4) we have that $\mathcal K(x,t)$ is a finite union of
points $v _i(x,t)$, thus we can evaluate the integral in terms of
\begin{equation*}
\int_{\mathcal K(x,t)}\sigma (\xi )\frac{d\xi }{|Df_{x,t}(\xi )|}
=\sum_{i=1}^{N}\frac{\sigma (v_i(x,t))}{|Df_{x,t}(v_i(x,t))|},
\end{equation*}
which in the sense of measures can be written in the form 
\begin{equation}
\label{dsum}\sum_{i=1}^{N}\frac{\left<\delta (\xi-v_i(x,t) ),\sigma(\xi
)\right>}{|Df_{x,t}(v_i(x,t))|}.
\end{equation}
The local differentiability of $v_i$, $i=1\dots N$, is a direct
consequence of the implicit function theorem. The existence of phase functions
$S_i(x,t)$ such that $v_i=\nabla _xS_i$ can be concluded essentially in the same
way as in Corollary 4.1, using 
\begin{equation}
f_{x, 0}(\xi )=\xi -\nabla _xS_I(x)
\end{equation}
and thus
\begin{equation}
f_{x,0}(v_i(x,0))=v_i(x,0) -\nabla _xS_I(x)=0,
\end{equation}
since by construction $v_i(x,t)\in\mathcal K(x,t)$. Thus we obtain
(\ref{multiphas}) by inserting (\ref{dsum}) into (\ref{dide}).
\end{proof}
\noindent In view of the above theorem we can now define define the
\emph{caustic set} $\mathcal C\subset \mathbb R_x^d\times \mathbb R_t$ by
\begin{definition}
\begin{equation} \label{Cdef}
\mathcal C:=\{(x,t): Df_{x,t}(\xi )=0 \mbox{ \emph{for at least one} } \xi \in
\mathcal K(x,t)\}
\end{equation} 
\end{definition}

\noindent This definition becomes more clear by the following lemma.
\begin{lemma} Denote by $\hat x^{-1}(t,x)\subseteq \mathbb R^d$ the pre-image of
the point $(x,t)$ under the flow $\hat x(t, \cdot )$ (i.e. the set of all points
$z$ with $\hat x(t,z )=x$) and by $\check x(t,\cdot )$ the (locally defined)
inverse map of $\hat x(t,\cdot )$. Let $\mathcal V\subseteq \mathbb R^d_x\times
\mathbb R_t$ be such that there exists only one $v=\nabla _xS\in\mathcal
K(x,t)$, then we have 
\begin{equation}
|Df_{x,t}(\nabla _xS)|\equiv \Jac (\check x(t,x),t),\quad \forall \
(x,t)\in\mathcal V
\end{equation}
where 
\begin{equation} \label{Jacobian1}
\Jac(z,t): =\vert \det \left(\frac{\partial \hat x(t,z)}{\partial
z}\right)\vert.
\end{equation} 
\end{lemma}
\noindent Thus \emph{the Jacobian $\Jac$ of the rays $\hat x(t,x)$ becomes zero
at the caustics}, c.f. theorem 4.3. 
This property is usually used to define caustics in the WKB-framework (see e.g.
\cite{BKM}, \cite{Fed}).
\begin{proof}
The claim follows directly from identity (\ref{ide}), having in mind that if
there exist only one $v\in\mathcal K(x,t)$ then $\hat x(t,\cdot )$ is locally
one-to-one and thus we can apply the transformation law of integrals (see
\cite{Fe}) on the right hand side of (\ref{ide}), to obtain
\begin{equation*}
n(x,t)=n_I(\check x(t,x)) \vert \det \left(\frac{\partial \check
x(t,x)}{\partial x}\right)\vert.
\end{equation*} 
Since $\hat x(t,\cdot )$ in this case is, by assumption, invertible the basic
calculus for determinants implies 
\begin{equation}
\label{n} n(x,t)=\frac{n_I(\check x(t,x))}{\Jac (\check x(t,x),t)}.
\end{equation}
Finally note that $\check x(t,x)=\tilde x(-t,x,v(x,t))$ with $v=\nabla _xS(t,x)$
as long as $\hat x(t,\cdot )$ is one-to-one. Thus the claim is proven.
\end{proof}
\noindent Note that (\ref{n}) is the solution of the conservation law (\ref{wkb1}) before the caustic onset, i.e. for 
$t_{c_1}<t<t_{c_2}$. \\
We have obtained a multivalued description of the solution of the
WKB-system for all $t\in \mathbb R$, locally away from caustics $\mathcal C$, by
representing the limiting phase space density $w(t)$ as a sum over mono-kinetic
terms, each of which can be associated to a single branch of the multivalued
solution of the Hamilton-Jacobi equation (\ref{wkb2}). Note however that, since 
$\nabla _xS_i =\nabla _x \mathcal S_i$ with $\mathcal S_i (x,t)=S_i (x,t)+C_i$, $C_i\in\mathbb R$, it is 
clear that for each $v_i\in\mathcal K$ we obtain the corresponding phase $S_i$ only up to a constant.\\

\noindent \textbf{Remark.} Physically the multivaluedness can be interpreted as
interference of the wave with itself, or in terms of classical mechanics, that
faster particles overtake slower ones for times $t \leq t_{c_1}$, $t\geq
t_{c_2}$.\\

\noindent To close the argument, the following corollary shows, that the Wigner measure given by (\ref{multiphas}) 
is  indeed the limiting measure of a superposition of (WKB) waves.
\begin{corollary} Assume (A1)-(A4). and let $\mathcal U \cap \mathcal C
=\emptyset$,
with $C$ defined in \fer{Cdef}. 
Define for all $(x,t)\in \mathcal U$ a generalized WKB-asymptotic solution by
\begin{equation} \label{WKB-N-rep}
\psi^\varepsilon _{wkb}(x,t):=
\sum_{k=1}^{N}\sqrt{n_i(x,t)}\exp\left(\frac{i}{\varepsilon }S_i(x,t)\right),
\quad (x,t)\in\mathcal U
\end{equation}
where $\nabla _xS_i\in\mathcal K$, $\forall $ $i=1\dots N$ and each branch of the energy density reads
\begin{equation} \label{nidef}
n_i(x,t):=\frac{n_I\left(\tilde x(-t,x,\nabla
_xS_i(x,t))\right)}{|Df_{x,t}(\nabla _xS_i(x,t))|}.
\end{equation}
For a localization 
let $\phi \in C^\infty _0(\mathbb R_x^d\times \mathbb R_t)$ with $\supp\phi 
\subseteq \mathcal U$. 
Then the unique (semi-)classical Wigner measure $w_\phi $ of 
$w^\varepsilon [\phi  \psi^\varepsilon _{wkb}]$ 
is given by (\ref{multiphas}) multiplied by $\phi ^2$.
\end{corollary}
\begin{proof} Having in mind lemma 3.1, the proof follows directly from the fact
(see \cite{Ge}) that if $f^\varepsilon$, $g^\varepsilon$ have Wigner measures
$w_f$, $w_g$, which are mutually singular, i.e. there exist two disjoint
Borel-sets $\mathcal A$, $\mathcal B$ with $w_f(\mathcal A^c)=0$, $w_g(\mathcal
B^c)=0$, then
\begin{equation}
w^{\varepsilon _k}[f^\varepsilon+g^\varepsilon ]\stackrel{\varepsilon
\rightarrow 0 }{\longrightarrow}w _f+w_g \mbox{ in $\mathcal S'(\mathbb
R_x^d\times \mathbb R_\xi ^d)$}.
\end{equation}
Clearly this is true in our case, since each term in the expression
(\ref{multiphas}) is a measure which is supported on $\{(x,\nabla
_xS_i(x,t))\}$. By construction, we have $\nabla_x S_i\not=\nabla_xS_j$ for
$i\not=j$, which implies the assertion.\\
\end{proof}
\noindent Observe that $\psi^\varepsilon _{wkb}$ 
\emph{does not explicitly contain the so called "Maslov-phase shifts"}
(c.f. section 2.3) due to the fact that each branch of the multivalued phase $S_i$, obtained by our
kinetic approach, is uniquely determined only up to a constant, which is not explicitly specified
by the approach.
In each region $\mathcal U$ the Maslov indices are such constants. 
More precisely, we connect the notation used in our Wigner measure approach with the one used in section 2.3 (FIO's) by 
stressing that 
\begin{align*}
z_i(x,t)&=\tilde x(-t,x,\nabla _xS_i(x,t)) ,\\
S_i(x,t)&=S_I(z_i(x,t))+\frac{\varepsilon \pi }{4}m_i.
\end{align*}
On the other hand it is clear, see section 2.3 and e.g. \cite{GuSt} for more details, that once the the Lagrangian manifold generated 
by $\nabla _xS_i(t,x)$ is given, the Maslov phase shift can be calculated from it from purely geometrial considerations. 
Thus although it does not appear explicitly, the Wigner measure (in the representation (\ref{multiphas})) contains all 
information needed to obtain the phase shift, which then allows the construction of the correct multivalued WKB-expansion after the caustic.\\

\noindent \textbf{Remark.} Although the Wigner measure gives the correct multivalued description after the first caustic its computation
in general is quite labourintensive, in particular since it involves $2d+1$ variables. From a numerical point of view 
one would like to work in physical space $\mathbb R^d_x\times \mathbb R_t$.\\ This requires the approximation 
of $w$ by a system of $2N$ equations for the $2N$ unknowns $(n_i,v_i)$, $1\leq i\leq N$. 
This system for the $2N$ moments of $w$ in general is not closed. However, in geometrical optics, the multivalued form 
of the Wigner measure (for a fixed $N$) gives a closing condition which allows (in principle) the correct description of multivalued situations 
until the next caustic forms. For details on this problem (in $d=1$) see \cite{JiLi} and for some alternative approaches we refer to 
\cite{Be1}, \cite{Be2}, \cite{Ru}.

\subsection{Concentration effects}

\noindent We now describe the behavior of the density at focal points, which
typically arise as the onset of caustics. We will be able to distinguish between
two specific cases of energy concentrations. \\
\begin{theorem} Fix $t\not =0, y\in\mathbb R^d$ and let $\mathcal V\subseteq
\mathbb R^d_x\times \mathbb R_t$ be a region (closed set), with
$(y,t)\in\mathcal V$, such that there exists only one $v\in\mathcal K(x,t)$ for
all $(x,t)\in \mathcal V$. Further assume that there exists $r>0$ such that
$\{(x,t): |y-x|<r\}\subseteq \mathcal V$. Then we have 
\begin{equation}
\label{concentr}n(x,t)= \frac{n_I\left(\tilde
x(-t,x,v)\right)}{|Df_{x,t}(v(x,t))|}\chi _{\{x\not=y\}} + \mu \delta
(x-y),\quad |y-x|<r
\end{equation}
where $\chi$ denotes the characteristic function and
\begin{equation}
\label{mass}\mu :=\int_{\{\hat x^{-1}(y,t)\}}n_I(z)dz.
\end{equation}
\end{theorem}
\begin{proof} Applying identity (\ref{ide}) to the monokinetic form of the
initial data with $\varphi (x,\xi )=\gamma (x)$ we obtain
\begin{equation*}
\int_{\mathbb R^d}\gamma (x) n(dx,t)=\int_{\mathbb{R}^d}n_I(x)\gamma  (\hat
x(t,x)) dx 
\end{equation*} 
Thus we can write 
\begin{eqnarray}
\label{conprof}\int_{\mathbb R^d}\gamma (x)
n(dx,t)&=&\int_{\mathbb{R}^d}n_I(x)\gamma  (\hat x(t,x)) \chi
_{\{x\not\in  \hat x^{-1}(y,t)\}}dx\nonumber\\
&+&\int_{\mathbb R^d}n_I(x)\gamma  (\hat x(t,x)) \chi _{\{x\in  \hat
x^{-1}(y,t)\}}dx.
\end{eqnarray} 
Now let $\gamma (\hat x(t,\cdot ))$ be supported in $\{x\in\mathbb
R^d:|y-x|<r\}$. In the last term of the right hand side we obtain
\begin{eqnarray*}
\gamma (y)\int_{\{\hat x^{-1}(y,t_c)\}}n_I(z)dz\equiv \left<\mu \delta
(x-y),\gamma (x)\right>
\end{eqnarray*} 
where $\left<\cdot ,\cdot \right>$ denotes the duality bracket. In the
first term on the right hand side the transformation law of integrals can be
applied thus using lemma 4.2 we obtain the expression (\ref{concentr}) which
proves the claim.\\
\end{proof}
\noindent If there is a nonzero amount of initial mass $\mu$ carried by rays
into the point $(y,t)$ the energy density $n$ \emph{"concentrates"} as
the last term of (\ref{concentr}) shows. This is also an explanation of the word
\emph{"caustic"}, since its Greek origin means \emph{"which
burns"}. We shall refer to these caustic points as
\emph{"hot"}.\\

\noindent We deduce from the theorem above the following easy consequences:
\begin{corollary} If  $w_I(\{\hat x^{-1}(t,y)\}\times \mathbb R_\xi ^d)=0$, then
locally around $y$
\begin{equation}
\label{concentr2}n(x,t)= \frac{n_I(\check x(t,x))}{\Jac (\check x(t,x))}\chi
_{\{x\not=y\}},\quad(x,t)\in\mathcal V.
\end{equation}
In particular the density remains in $L^1_{loc}(\mathbb R^d)$ in this case.
Clearly if $\hat x(t,\cdot )$ is a diffeomorphism in $\mathcal V$ the expression
(\ref{concentr2}) is valid for all $(x,t)\in \mathcal V$.
\end{corollary}
\begin{proof} The claim follows from $n_I(\mathcal A)=0$ iff $w_I(\mathcal
A\times \mathbb R_\xi ^d)=0$, where $\mathcal A$ is an arbitrary Borel set.
\end{proof}
\noindent It is important to note, that although the amplitude blows up at the caustic $\mathcal C$, the density
$n$, still makes sense as a measure. For this reason the amplitude is sometimes called 
\emph{half-density} \cite{Ho}.\\
In other words: Although some rays $\hat x(t,\cdot )$ may cross, the density, as
the corollary shows, may still be in $L^1_{loc}(\mathbb R^d)$. This will be referred to as a
\emph{"cool"} caustic point, in contrast to a hot caustic where we
obtain a concentration of the density. A particular case of such a cool caustic
is given in example 1.3. in section 5 below. (See also \cite{MPP} for a
numerical study.) Another cool caustic is given by example 1 in \cite{GaMa}. \\

\noindent \textbf{Remark.} Clearly the above theorem and corollary can be extended to regions in
which finitely many (gradients of) phases appear. This is in particularly true
for points on the one dimensional cusp-caustic (see example 1.3), where within
the caustic region we have 3 zeros $v_i(x,t)$ and outside there is only one.
Thus we obtain cool focus points on each branch of the caustic, whereas the
caustic-onset point, or \emph{focus point}, is hot.\\


\section{Case studies}
We now illustrate the above analysis with examples. \\

\noindent\textbf{1. Free motion}\\

\noindent One should note that the following calculations are generalizations of the 
formulas given in the example of the free Schr\"odinger equation in section 2.3.\\

\noindent Although the  generalized free motion case, i.e. 
\begin{equation}
H=H(\xi ),
\end{equation}
is the most simple one, it nevertheless features a remarkable variety of
interesting phenomena. 
The associated Hamiltonian flow is given by
\begin{equation}
\label{free}F_t(x,\xi)= (x+t\nabla _\xi H(\xi),\xi ).
\end{equation}
The velocity remains constant $\tilde \xi(t,x,\xi)=\xi$, $\forall t\in\mathbb R,
x,\xi \in\mathbb R^d$ and thus the rays are straight lines, with slope $\nabla
_\xi H(\nabla _xS_I(x))$, i.e. we have $\hat x(t,x)=x+t\nabla _\xi H(\nabla
_xS_I(x))$. In this particular case definition 4.2 reads 
\begin{equation}
f_{x,t}(\xi ):=\xi -\nabla _xS_I(x-t\nabla _\xi H(\xi) )
\end{equation}
and its zeros $v_i(x,t)\in \mathcal K(x,t)$ satisfy the implicit relation
\begin{equation}
v_i(x,t)=\nabla _xS_I(x-t \nabla _\xi H(v_i(x,t))).
\end{equation}
This is the well known (multivalued) implicit solution formula of the
conservation law
\begin{align}
\partial _tv + \nabla_x H(v)&=0,\quad x\in \mathbb R^d, t\in \mathbb R \\
v _I(x,0)&=\nabla _xS_I(x),
\end{align}
which holds as long as the determinant of the Jacobian is nonzero
\begin{equation}
\label{jaco}Df_{x,t}(v_i)= \det(I+t D^2S_I(x-t\nabla _\xi H(v_i))
D^2H(v_i))\not=0
\end{equation}
(Here $D^2f$ denotes the Hessian of $f$.) The multivalued phase $S_i(x,t)$ is
obtained, locally in each region $\mathcal U$ in which $N$ is constant, by using
standard Hamilton-Jacobi theory, i.e. by integrating (\ref{phas}) 
\begin{equation*}
\frac{dS_i(\hat x,t)}{dt}=H'(\hat \xi )\cdot \hat \xi -H(\hat \xi ), \quad
S_i(x,0)=S_I(x)
\end{equation*}
after inserting $\hat \xi (t,x)=v_i(\hat x(t,x),t)$ and using $v_i(x,0)=\nabla
_xS_I(x,0)$ to determine the initial condition.\\
Finally, it follows from (\ref{multiphas}), that for all $(x,t)\in\mathcal U$,
$(x,t)\not\in\mathcal C$, the density is given by
\begin{equation}
\label{mdens}n(x,t)=\sum_{i=1}^{N}\frac{n_I(x-t \nabla _\xi H(v_i))}{|\det(1+t
D^2S_I(x-t\nabla _\xi H(v_i)) D^2H(v_i))|}.
\end{equation}
\noindent In the examples 1.1.-1.3. below we restrict ourselves, for
simplicity to the case of one spatial dimension, i.e. $d=1$. We further assume
that $H(\xi )$ is equal to the classical kinetic energy 
\begin{equation}
H(\xi )=\frac{\xi ^2}{2},\quad\xi \in\mathbb R,
\end{equation}
which corresponds to case of the free Schr\"odinger equation. 
Then we analyse for $t\geq 0$ the behavior of the WKB-system subject to
different types of initial phases $S_I$.\\

\noindent \textbf{Example 1.1.}  \emph{"No caustic"}
\begin{equation}
S_I(x):=\frac{x^2}{2},\quad x \in\mathbb R  
\end{equation}
Here the rays $\hat x(t,x)$ never cross, instead they spread out, forming a so
called \emph{rarefaction wave}(see fig. 1). The only element $v(x,t)\in \mathcal
K(x,t)$ is given by
\begin{equation}
v(x,t)=\frac{x}{t+1}.
\end{equation}
The solution of the Hamilton-Jacobi equation, which is single valued and smooth
for all times $t\geq 0$, reads
\begin{equation}
S(x,t)=\frac{x^2}{2(t+1)} 
\end{equation}
and the limiting density (\ref{mdens}) simplifies to
\begin{equation}
n(x,t)=\frac{1}{|t+1|}\ n_I\left(\frac{x}{t+1}\right).
\end{equation}
\

\noindent \textbf{Example 1.2.} \emph{"Focusing at a point"}\\

\noindent By simply changing the sign of the initial phase we obtain
\begin{equation}
S_I(x):=-\frac{x^2}{2},\quad x \in\mathbb R,
\end{equation}
which leads to the \emph{single focus case}. All rays intersect at the hot focus
point $(x,t_c)=(0,1)$ and spread afterwards (see fig. 2), i.e. there is a.e.
only one phase
\begin{equation}
S(x,t)=\frac{x^2}{2(t-1)},\quad t\not=1.
\end{equation}
The above theory does not tell us anything about the precise description of the
phase $S(x,t)$ at $t=1$. With the use of the theorem 4.2 and equation
(\ref{mdens}) the density is given by
\begin{eqnarray}
n(x,t)&=&\frac{1}{|t-1|}\ n_I\left(\frac{x}{t-1}\right)\quad t\not=1,\\
n(x,1)&=&\int_{\mathbb R}n_I(x)dx\ \delta (x).
\end{eqnarray}
In this example there exists a generalized solution $(n,\sigma (v))$ of
(\ref{renorm}) $n(t)$-a.e..\\

\noindent \textbf{Example 1.3.} \emph{"Caustic"}\\

\noindent We choose
\begin{equation}
S_I(x):=-\ln(\cosh(x)),\quad x \in\mathbb R
\end{equation}
such that the initial data for the ODE-system of rays is
"\emph{compressing}":
\begin{equation}
S_I'(x):=-\tanh(x).
\end{equation}
The equation which characterizes the kernel $\mathcal K(x,t)$ cannot be solved
explicitly, however a precise numerical study is given in \cite{MPP}. We want to
stress again that in this case no "hot" focus appears, i.e. the
limiting density $n\in L^1_{loc}(\mathbb R)$ for all $t\in\mathbb R$.\\
An explicitly solvable example (see also \cite{GaMa}) which has a similar
qualitative behavior (except that the focus is hot) is given by
\begin{equation}
S_I(x):=x\chi _{\{x<0\}}+(x-\frac{x^2}{2})\chi _{\{0\leq x\leq 1\}},\quad x
\in\mathbb R
\end{equation}
where $\chi $ denotes the characteristic function. Note that the initial
condition is only Lipschitz continuous, i.e. differentiable almost
everywhere.\\
\noindent Up to the time $t=1$ the rays do not intersect, at $t=1$ a focus point
occurs at $(x,t_c)=(1,1)$ from which two caustics emanate, forming a \emph{cusp}
(see fig. 3). By applying theorem 4.2, in a neighborhood of the focus, we obtain
for the density
\begin{equation}
n(x,1)=n_I(x)\Theta (x-1)+n_I(x-1)\Theta (1-x)+ \mu \delta (x-1).
\end{equation}
Here the amount of mass that gets concentrated at the point $(1,1)$ is given by
$\mu :=\int_0^1n_I(x)dx $ and $\Theta(x)$ denotes the Heaviside function.\\
For $t>1$ the solution of the Hamilton-Jacobi equation is triple-valued (as
it is generic for the singularities in one dimension) within the region
$1<x\leq t$, since $f_{x,t}(\cdot )$ has three zeros there
\begin{equation}
v_1(x,t)=1,\quad v_2(x,t)=\frac{1-x}{1-t},\quad v_3(x,t)=0
\end{equation}
and thus we get for the density within the caustics
\begin{equation}
n(x,t)\chi_{(1<x\leq
t)}=n_I(x-t)+\frac{1}{|t-1|}n_I\left(\frac{x-t}{1-t}\right)+n_I(x).
\end{equation}
The corresponding phases are obtained by a simple integration using
(\ref{phas}), for example the phase function, that gets transported into the caustic region from the left, 
is given by
\begin{equation}
S_1(x,t)=x-\frac{t}{2}.
\end{equation}

\noindent We now turn to an explicitly solvable case with $x$-dependent
Hamiltonian $H(x,\xi )$.\\

\noindent\textbf{2. Harmonic Oscillator}\\

\noindent Consider in the example of the harmonic oscillator (see also
\cite{GaMa})
\begin{equation} 
H(x,\xi )=\frac{|\xi|^2}{2}+\frac{|x|^2}{2}\quad x,\xi \in \mathbb R^d
\end{equation}
In this case the Hamiltonian flow $F_t$ is given by
\begin{eqnarray*}
\tilde x(t,x,\xi )&=&x\cos t +\xi \sin t\\
\tilde \xi (t,x,\xi )&=&-x\sin t +\xi \cos t.
\end{eqnarray*}
If we choose in particular 
\begin{equation}
S_I(x):=kx,\quad k>0,x \in\mathbb R
\end{equation}
we obtain a constant initial velocity $v_I(x)=k>0$, such that all rays
intersect at hot focal points (see fig. 4) given by
\begin{equation}
(x,t_c)=\left((-1)^{m+1}k, (2m+1)\pi/2 \right)\quad m\in \mathbb{Z}.
\end{equation}
\noindent We obtain a.e. only one zero $\xi =v(x,t)\in\mathcal K(x,t)$ of
$f_{x,t}(\xi )$, given by 
\begin{equation*}
v(x,t)=\frac{k-x\sin t}{ \cos t  },\quad t\not=(2m+1)\frac{\pi }{2}
\end{equation*}
and thus the solution of the Hamilton Jacobi equation is unique $n(t)$-a.e. 
\begin{equation}
S(x,t)=-\frac{1}{2}(x^2+k^2)\tan t + \frac{kx}{\cos t},\quad
t\not=(2m+1)\frac{\pi }{2}.
\end{equation}
The density is given by
\begin{eqnarray}
n(x,t)&=&\frac{1}{\vert \cos t\vert^d} \ n_I\left(\frac{x-k\sin t}{\cos
t}\right),\quad t\not=(2m+1)\frac{\pi }{2}, \\
n(x,t)&=&\int_{\mathbb R^d} n_I(z)dz \ \delta (x-(-1)^mk),\quad
t=(2m+1)\frac{\pi }{2}. 
\end{eqnarray}
In this case again there exists a generalized solution $(n,\sigma (v))$ of
(\ref{renorm}) $n(t)$-a.e..\\


\section{Conclusion}

In this paper we compare two important approaches within the theory of geometrical optics for linear 
dispersive operators (acting on $\psi ^\varepsilon $), namely the Wigner transform and time-dependent WKB-asymptotics. 
Whereas the latter faces the problem of caustics the limiting Wigner measure is insensitive for such 
obstacles, due two the fact that it lives in a $2d$-dimensional phase space in which the appearing singularities become unfolded. 
This feature makes the Wigner transform the method of choice if one is particularly interested in the high frequency behavior of energy 
densities, which are obtained as moments (w.r.t. the velocity variable $\xi $) of the Wigner measure. \\
Maybe the most important conclusion from this paper is the fact that for WKB-initial data the limiting Wigner measure $w$ can be (locally away
from caustics) decomposed into a sum of monokinetic terms (\ref{multiphas}). Each term carries all the information needed, namely $(n_i, S_i)$, $i=1\dots N$,  
to obtain a (local) approximation $\psi ^\varepsilon _{wkb}$ of the solution $\psi ^\varepsilon $ to the dispersive equation at the order $O(\varepsilon )$, 
despite the fact that the measure $w$ does not explicitly contain the phase shifts obtained by the stationary phase method used in the FIO approaches.
However since it is well known that, given $S_i$, these phase shifts can be obtained by a purely geometrical computation, we conclude that 
these important feature of geometrical optics is actually hidden in $w$.\\
In other  words the Wigner measure which is a description for $\varepsilon =0$ does indeed allow to obtain a 
approximation of $\psi^\varepsilon $ for finite $\varepsilon $ and can thus be seen as a necessary 
(but of course not sufficient) condition for the validity of the multivalued WKB approximation.
Higher order correction can be also obtained from the Wigner transform (using the expansion (\ref{exp})), although 
we neglected them in this work for the sake of simplicity. \\
We further stress the fact that although the amplitude of the WKB approach becomes infinite 
at the caustic, the energy density $n$ ("square of the amplitude") remains well defined in the sense of measures, as theorem 4.3 shows. 
What can not be obtained from the Wigner measure is information on the phase at caustic points, which is clearly in agreement with the 
fact that the multivalued WKB approximation breaks down there too and the fact that the only correct description 
at caustics is given by FIO's.\\  
It remains to say that the Wigner transform can be used in situation with much less regularity of the initial data than the WKB approach 
or its generalizations, since all our computations remain valid for $n_I\in L^1$ and $S_I \in C^1$ with Lipshitz continuous derivative.\\

\vskip 1cm 
\noindent\textbf{Acknowledgment}\\

\noindent We want to thank Y. Brenier and R. Seiringer for helpful
discussions and especially C. Bardos for valuable criticism that significantly shaped
the final version of this work for the benefit of the reader.\\


\section{Appendix}

As an appendix to the main results of this paper we first present a (rather
exotic) example where key assumptions of the presented theory are violated. Then we briefly comment on the 
physical interpretation of the fluid type equations, which arise in our work.\\


\subsection{Appendix 1 : A counterexample to global Hamiltonian flow}

As a "caveat" that the assumption of essential self adjointness (A1)
(ii) and global Hamiltonian flow  (A3)
are not trivial we consider a variable coefficient Airy-type equation \fer{airy}
in one space dimension
\begin{equation} \label{Hairy}
H(x,\xi)=-x\xi^3,\quad x,\xi \in\R.
\end{equation}
A lengthy calculation shows that the Weyl-quantized Hamiltonian operator 
corresponding to this symbol does not have a self adjoint extension. 
\noindent Moreover the Hamiltonian flow $F_t$ associated to \fer{Hairy}
is only locally defined as an explicit calculation shows:
\begin{eqnarray}
\tilde{x}(t,x,\xi)& = & x(1-2\xi^2t)^{3/2} \label{airyflow1}, \\
\tilde{\xi}(t,x,\xi)& = & \frac{\xi}{\sqrt{1-2\xi^2t}} \: , \quad
t<\frac{1}{2\xi^2}. 
\label{airyflow2}
\end{eqnarray}
\\
We remark that to our knowledge it is an open
question if essential self 
adjointness of $H^W(.,\eps D)$ is a sufficient condition for global Hamiltonian
flow defined by $H(x,\xi)$. \\
If we again choose
\begin{equation}
S_I(x):=kx \: , \quad k>0 ~, \: x\in\R,
\end{equation}
all rays $\hat x(t,x)$ focus at the point $(x,t_c)=(0,1/2k^2)$, i.e. we again
obtain a single focus case as in example 1.3. above. The corresponding phase-function is
given by
\begin{equation}
S(x,t)=\frac{kx}{\sqrt{1-2k^2t}} \: , \quad t<\frac{1}{2k^2}.
\end{equation}
It is not possible to extend the solution beyond the break time  since
$F_t(x,\xi =k)$ is only defined for $t<t_c$. \\
In this example neither the WKB nor the Wigner approach give an asymptotic
description of solutions of \fer{disp}, \fer{ho} after breaktimes.\\
We further remark that changing the sign in the Hamiltonian function gives rise
to an
expansive flow of rarefaction-wave type.

\subsection{Appendix 2: On the arising fluid-type equations}

As indicated above the fluid-type system (\ref{euler1}), (\ref{euler2}) admits a
physical interpretation if the Hamiltonian function is given by $H=|\xi
|^2/2+V(x)$ (see the remark below theorem 4.1). In general this is not the case.\\
However we state in the lemma below that one can find indeed an equivalent
system which is of the same form as the Euler-equation of gas dynamics, and
which can be defined for general Hamiltonian functions. \\

\noindent We define the generalized velocity $u$ and the (time dependent) modified force term $f$ by
\begin{align}
\label{gvel}u(x,t) &:= \nabla _\xi H(x,v(x,t)),\\
\label{gforce}f(x,t)&:= \{\nabla _\xi H,H \}\big|_{(x,v(x,t))},
\end{align}
where $v(x,t)$ is a $d$-dimensional vector field. 
With this definitions we obtain the following result:
\begin{lemma}
Let $(n,v)$ be a smooth solution of (\ref{euler1}), (\ref{euler2}) and let $u,f$
be defined as above on the same time interval $(t_{c_1},t_{c_2})$, then $(n,u)$
satisfies
\begin{eqnarray}
\label{meuler1}\partial _tn+\diverg(nu)=0,\quad n(x,0)=n_I(x)\\
\label{meuler2}\partial _t(nu)+\diverg(nu\otimes u)+nf=0,\quad u(x,0)=u_I(x)
\end{eqnarray}
with $u_I(x):=\nabla _\xi H(x,v_I(x))$.
\end{lemma}
\begin{proof} By definition (\ref{gvel}), it is clear that the conservation law
(\ref{meuler1}) is equivalent to (\ref{euler1}). Further note, that after
eliminating $n(x,t)$ from (\ref{meuler2}) using (\ref{meuler1}) one obtains for
$u(x,t)$ the \emph{Burgers equation} with source term 
\begin{equation}
\label{fdburg}\partial _tu+(u\cdot \nabla _x) u+f =0.
\end{equation}
Thus it remains to show that if $u(x,t)$ solves (\ref{fdburg}), then $v(x,t)$ is
a solution of 
\begin{equation}
\label{btype}\partial _tv+(\nabla_\xi H(x,v)\cdot \nabla _x) v+\nabla _x H(x,v )
=0,
\end{equation}
which is again obtained from (\ref{euler2})  by elimination of $n(x,t)$.
Calculating inner derivatives we obtain for the $i$-th component of
(\ref{fdburg}) 
\begin{equation*}
\sum_{k=1}^d[\partial _{\xi _k}(\partial _{\xi _i}H)][\partial
_tv_k+\sum_{l=1}^d\partial _{\xi _l}H\partial _{x_l}v_k]+\sum_{l=1}^d(\partial
_{\xi _l}H)\partial _{y_l}(\partial _{\xi _i}H)+f_i=0.
\end{equation*}
In order to have a classical solution $v(x,t)$ of (\ref{btype}) this implies
that each component  $f_i$, $i=1\dots d$, of the modified force term must
satisfy the relation
\begin{equation*}
f_i(x,t)=\sum_{k=1}^d(\partial^2 _{\xi _k\xi _i}H)\partial
_{y_k}H-\sum_{l=1}^d(\partial _{\xi _l}H)\partial _{y_l}(\partial _{\xi _i}H)
\end{equation*}
which is exactly definition (\ref{gforce}) above and the claim is proved.
\end{proof}
\noindent Clearly the fluid-type system formulated in $(n,v)$ and the Euler system in
$(n,u)$ are equivalent (for smooth solutions) if the relation (\ref{gvel}) can
be uniquely solved for $v$ in terms of $u$. In case 
\begin{equation}
H= \omega (\xi )+V(x),
\end{equation}
where $\omega $ is a general dispersion relation, we obtain $u=\nabla _\xi
\omega (v)$ and $f=D^2\omega (v)\nabla _xV$ (where again $D^2\omega $ denotes
the Hessian matrix of $\omega $) and thus
\begin{eqnarray}
\partial _tn+\diverg(nu)&=&0, \quad n(x,0)=n_I\\
\partial _t(nu)+\diverg(nu\otimes u)+nD^2\omega (v)\nabla _xV&=&0,\quad
u(x,0)=u_I.
\end{eqnarray}
Further note that by definition $u$ in general is not the gradient of a phase function $S$.




\end{document}